\documentclass{article}

\usepackage{arxiv}

\usepackage[utf8]{inputenc} 
\usepackage[T1]{fontenc}    
\usepackage{hyperref}       
\usepackage{url}            
\usepackage{booktabs}       
\usepackage{amsfonts}       
\usepackage{nicefrac}       
\usepackage{microtype}      
\usepackage{lipsum}
\usepackage{hyperref}
\usepackage{todonotes}
\graphicspath{{figures/}}
    \DeclareGraphicsExtensions{.pdf,.eps,.png,.jpg}
\usepackage[noabbrev,capitalise]{cleveref} 
	\crefname{equation}{}{}	

\usepackage[nolist]{acronym}
  \begin{acronym}[DRAM]
    \acro{AI}{artificial intelligence}
    \acro{ANN}{artificial neural network}
    \acrodefindefinite{AI}{an}{an}
    \acro{ASIC}{application-specific integrated circuit}
    \acrodefindefinite{ASIC}{an}{an}
    \acro{BD}{bounded delay}
    \acro{BC}{breast cancer}
    \acro{BNN}{binarized neural network}
    \acro{CB}{connectionist bench}
    \acro{CD}{completion detection}
    \acro{CNN}{convolutional neural network}
    \acro{CTM}{convolutional Tsetlin machine}
    \acro{CPOG}{conditional partial order graph}
    \acro{DI}{delay insensitive}
    \acro{DR}{dual-rail}
    \acro{DRAM}{dynamic RAM}
    \acro{DSP}{digital signal processing}
    \acro{DVFS}{dynamic voltage and frequency scaling}
    \acro{FD-SOI}{fully-depleted silicon-on-insulator}
    \acro{FPGA}{field-programmable gate array}
    \acro{FSM}{finite state machine}
    \acrodefindefinite{FPGA}{an}{a}
    \acro{HDL}{hardware description language}
    \acrodefindefinite{HDL}{an}{a}
    \acro{HVR}{house voting record}
    \acro{INWE}{inverse-narrow-width effect}
    \acrodefindefinite{INWE}{an}{an}
    \acro{ISA}{instruction set architecture}
    \acrodefindefinite{ISA}{an}{an}
    \acro{IoT}{Internet of Things}
    \acrodefindefinite{IoT}{an}{an}
    \acro{LA}{learning automaton}
    \acrodefplural{LA}{learning automata}
    \acrodefindefinite{LA}{an}{a}
    \acro{LEC}{logical equivalence checking}
    \acrodefplural{LEC}{logical equivalence checks}
    \acro{LFSR}{linear feedback shift register}
    \acrodefindefinite{LFSR}{an}{a}
    \acro{LU}{learning unit}
    \acrodefindefinite{LU}{an}{a}
    \acro{MAC}{multiply-accumulate}
    \acro{MNIST}{Modified National Institute of Standards and Technology }
    \acro{MEP}{minimum energy point}
    \acrodefindefinite{MEP}{an}{a}
    \acro{ML}{machine learning}
    \acrodefindefinite{ML}{an}{a}
    \acro{MLP}{multi-layer perceptron}
    \acrodefindefinite{MLP}{an}{a}
    \acro{MPP}{maximum power point}
    \acrodefindefinite{MPP}{an}{a}
    \acro{MPPT}{maximum power point tracking}
    \acrodefindefinite{MPPT}{an}{a}
    \acro{MRAM}{}
    \acro{NCL}{null convention logic}
    \acrodefindefinite{NCL}{an}{a}
    \acro{NN}{neural network}
    \acrodefindefinite{NN}{an}{a}
    \acro{OCV}{on-chip variation}
    \acrodefindefinite{OCV}{an}{an}
    \acro{PVT}{process, variation and temperature}
    \acro{QDI}{quasi delay insensitive}
    \acro{RAM}{random-access memory}
    \acrodefplural{RAM}{random-access memories}
    \acro{RDF}{random dopant fluctuation}
    \acro{RTM}{recurrent Tsetlin machine}
    \acrodefindefinite{RTM}{an}{a}
    \acro{SCM}{standard cell memory}
    \acrodefindefinite{SCM}{an}{a}
    \acrodefplural{SCM}{standard cell memories}
    \acro{SI}{speed independent}
    \acrodefindefinite{SI}{an}{a}
    \acro{SR}{single-rail}
    \acrodefindefinite{SR}{an}{a}
    \acro{SRAM}{static RAM}
    \acrodefindefinite{SRAM}{an}{a}
    \acro{STG}{signal transition graph}
    \acrodefindefinite{STG}{an}{a}
    \acro{SVM}{support vector machine}
    \acrodefindefinite{SVM}{an}{a}
    \acro{TA}{Tsetlin Automaton}
    \acrodefplural{TA}{Tsetlin Automata}
    \acro{TAT}{Tsetlin automaton team}
    \acro{TM}{Tsetlin machine}
    \acro{ULV}{ultra-low voltage}
    \acrodefindefinite{ULV}{a}{an}
    \acro{VLSI}{very-large-scale integration}
    \acro{WSN}{wide sensor network}
    \acro{IoT}{Internet of Things}
    \acro{IC}{integrated circuit}
    
    \acro{MFCC}{Mel-frequency cepstrum coefficients}
    \acro{KWS}{keyword spotting}
    
    \acro{DNN}{Deep Neural Network}
    \acro{QCNN}{Quantized Convolutional Neural Network}
    \acro{BCNN}{Binarized Convolutional Neural Network}
    \acro{LSTM}{Long Short-term Memory}
    \acro{SoC}{system on a chip}
    \acro{GPU}{Graphics Processing Unit}
    \acro{HPC}{High Performance Computing}
    \acro{CUDA}{Compute Unified Device Architecture}
  \end{acronym}

\title{Low-Power Audio Keyword Spotting \\ using Tsetlin Machines}

\author{
Jie Lei   \\
Microsystems Research Group \\
School of Engineering, \\ Newcastle University, NE1 7RU, UK \\
\textit{Jie.Lei@newcastle.ac.uk }\\
\and
Tousif Rahman  \\
Microsystems Research Group \\
School of Engineering,\\ Newcastle University, NE1 7RU, UK \\
\textit{S.Rahman@newcastle.ac.uk}\\
\and
Rishad Shafik  \\
Microsystems Research Group \\
School of Engineering,\\ Newcastle University, NE1 7RU, UK \\
\textit{Rishad.Shafik@newcastle.ac.uk }\\
\and
Adrian Wheeldon  \\
Microsystems Research Group \\
School of Engineering, \\ Newcastle University, NE1 7RU, UK \\
\textit{Adrian.Wheeldon@newcastle.ac.uk}\\
\and
Alex Yakovlev  \\
Microsystems Research Group \\
School of Engineering,\\ Newcastle University, NE1 7RU, UK \\
\textit{Alex.Yakovlev@newcastle.ac.uk }\\
\and
Ole-Christoffer Granmo  \\
Centre for AI Research (CAIR), \\ University of Agder, Kristiansand, Norway; \\
\textit{ole.granmo@uia.no} \\
\and
Fahim Kawsar  \\ 
Pervasive Systems Centre, \\ Nokia Bell Labs Cambridge, UK; \\
\textit{fahim.kawsar@nokia-bell-labs.com} \\
\and
Akhil Mathur   \\
Pervasive Systems Centre, \\ Nokia Bell Labs Cambridge, UK; \\
\textit{akhil.mathur@nokia-bell-labs.com} } 

\begin{document}
\maketitle

\begin{abstract}
The emergence of Artificial Intelligence (AI) driven Keyword Spotting (KWS) technologies has revolutionized human to machine interaction. Yet, the challenge of end-to-end energy efficiency, memory footprint and system complexity of current Neural Network (NN) powered AI-KWS pipelines has remained ever present. This paper evaluates KWS utilizing a learning automata powered machine learning algorithm called the Tsetlin Machine (TM). Through significant reduction in parameter requirements and choosing logic over arithmetic based processing, the TM offers new opportunities for low-power KWS while maintaining high learning efficacy. In this paper we explore a TM based \ac{KWS} pipeline to demonstrate low complexity with faster rate of convergence compared to NNs. Further, we investigate the scalability with increasing keywords and explore the potential for enabling low-power on-chip KWS.
\end{abstract}



\section{Introduction}

Continued advances in \ac{IoT} and embedded system design have allowed for accelerated progress in \ac{AI} based applications \cite{8789967}. \ac{AI} driven technologies utilizing sensory data have already had a profoundly beneficial impact to society, including those in personalized medical care~\cite{ML_healthcare}, intelligent wearables~\cite{Wearables} as well as disaster prevention and disease control~\cite{9308291}.

A major aspect of widespread~\ac{AI} integration into modern living is underpinned by the ability to bridge the human-machine interface, \textit{viz}. through sound recognition. Current advances in sound classification have allowed for AI to be incorporated into self-driving cars, home assistant devices and aiding those with vision and hearing impairments~\cite{8986304}. One of the core concepts that has allowed for these applications is through using \ac{KWS}\cite{BENISTY20181}. Selection of specifically chosen key words narrows the training data volume thereby allowing the \ac{AI} to have a more focused functionality \cite{Giraldo2019}. 

With the given keywords, modern keyword detection based applications are usually reliant on responsive real-time results~\cite{8777994} and as such, the practicality of transitioning keyword recognition based machine learning to wearable, and other smart devices is still dominated by the challenges of \textit{algorithmic complexity} of the \ac{KWS} pipeline, \textit{energy efficiency} of the target device and the \ac{AI} model's \textit{learning efficacy}.  

The \textit{algorithmic complexity} of \ac{KWS} stems from the pre-processing requirements of speech activity detection, noise reduction, and subsequent signal processing for audio feature extraction, gradually increasing application and system latency \cite{Giraldo2019}. When considering on-chip processing, the issue of algorithmic complexity driving operational latency may still be inherently present in the \ac{AI} model \cite{Giraldo2019, 8641328}.

\ac{AI} based speech recognition often offload computation to a cloud service. However, ensuring real-time responses from such a service requires constant network availability and offers poor return on end-to-end energy efficiency \cite{Sorensen2020}. Dependency on cloud services also leads to issues involving data reliability and more increasingly, user data privacy~\cite{Merenda2020}. 

Currently most commonly used \ac{AI} methods apply a \ac{NN} based architecture or some derivative of it in \ac{KWS} \cite{8641328,8936893,8777994,8502309} (see Section \cref{sec:RelatedKWS} for a relevant review). The \ac{NN} based models employ arithmetically intensive gradient descent computations for fine-tuning feature weights. The adjustment of these weights require a large number of system-wide parameters, called hyperparameters, to balance the dichotomy between performance and accuracy~\cite{Bacanin2020}. Given that these components, as well as their complex controls are intrinsic to the \ac{NN} model, energy efficiency has remained challenging~\cite{shafik2018real}. 

To enable alternative avenues toward real-time  energy efficient \ac{KWS}, low-complexity \ac{ML} solutions should be explored. A different ML model will eliminate the need to focus on issues \ac{NN} designers currently face such as optimizing arithmetic operations or automating hyper-parameter searches. In doing so, new methodologies can be evaluated against the essential application requirements of energy efficiency and learning efficacy,

The challenge of \textit{energy efficiency} is often tackled through intelligent hardware-software co-design techniques or a highly customized \ac{AI} accelerator, the principal goal being to exploit the available resources as much as possible.  

To obtain adequate \textit{learning efficacy} for keyword recognition the \ac{KWS}-\ac{AI} pipeline must be tuned to adapt to speech speed and irregularities, but most crucially it must be able to extract the significant features of the keyword from the time-domain to avoid redundancies that lead to increased latency.   

Overall, to effectively transition keyword detection to miniature form factor devices, there must be a conscious design effort in minimizing the latency of the \ac{KWS}-\ac{AI} pipeline through algorithmic optimizations and exploration of alternative \ac{AI} models, development of dedicated hardware accelerators to minimize power consumption, and understanding the relationships between specific audio features with their associated keyword and how they impact learning accuracy.  

This paper establishes an analytical and experimental methodology for addressing the design challenges mentioned above. A new automata based learning method called the \ac{TM} is evaluated in the \ac{KWS}-\ac{AI} design in place of the traditional perceptron based \ac{NN}s. The \ac{TM} operates through deriving propositional logic that describes the input features \cite{Granmo2018}. It has shown great potential over \ac{NN} based models in delivering energy frugal \ac{AI} application while maintaining faster convergence and high learning efficacy \cite{shafikexplainability,Wheeldon2020a,jieleitmnn}

Through exploring design optimizations utilizing the \ac{TM} in the \ac{KWS}-\ac{AI} pipeline we address the following research questions:
\begin{itemize}
    \item How effective is the \ac{TM} at solving real-world \ac{KWS} problems?
    \item Does the Tsetlin Machine scale well as the \ac{KWS} problem size is increased?
    \item How robust is the Tsetlin Machine in the \ac{KWS}-\ac{AI} pipeline when dealing with dataset irregularities and overlapping features?
\end{itemize}
This initial design exploration will uncover the relationships concerning how the Tsetlin Machine's parameters affect the \ac{KWS} performance, thus enabling further research into energy efficient \ac{KWS}-\ac{TM} methodology.
\subsection{Contributions}
The \textit{contributions} of this paper are as follows:
\begin{itemize}
    \item Development of a pipeline for \ac{KWS} using the \ac{TM}. 
    \item Using data encoding techniques to control feature granularity in the \ac{TM} pipeline.
    \item Exploration of how the Tsetlin Machine's parameters and architectural components can be adjusted to deliver better performance.
\end{itemize}

\subsection{Paper Structure}
The rest of the paper is organized as follows:
\cref{sec:wsp_design_tm} offers an introduction to the core building blocks and hyper-parameters of the Tsetlin Machine.
Through exploring the methods of feature extraction and encoding process blocks, the \ac{KWS}-\ac{TM} pipeline is proposed in \cref{sec:explainMFCCtmPipe}.
We then analyze the effects of manipulating the pipeline hyper-parameters in Section 4 showing the Experimental Results. We examine the effects of changing the number of \acp{MFCC} generated, the granularity of the encoding and the the robustness of the pipeline through the impact of acoustically similar keywords. We then apply our understanding of the Tsetlin Machines attributes to optimize performance and energy expenditure through \cref{sec:sweepClause}. 

Through the related works presented in \cref{sec:RelatedTM}, we explore the current research progress on \ac{AI} powered audio recognition applications and offer an in-depth look at the key component functions of the \ac{TM}.  
We summarize the major findings in \cref{sec:conclusions} and present the direction of future work in \cref{sec:future}.

\section{A Brief Introduction to Tsetlin Machine}\label{sec:wsp_design_tm}

The Tsetlin Machine is a promising, new \ac{ML} algorithm based on formulation of propositional logic~\cite{Granmo2018}. This section offers a high level overview of the main functional blocks; a detailed review of relevant research progress can be found in \cref{sec:RelatedTM}. 

\begin{figure}[htbp]
    \centering
    \includegraphics[width=0.7\columnwidth]{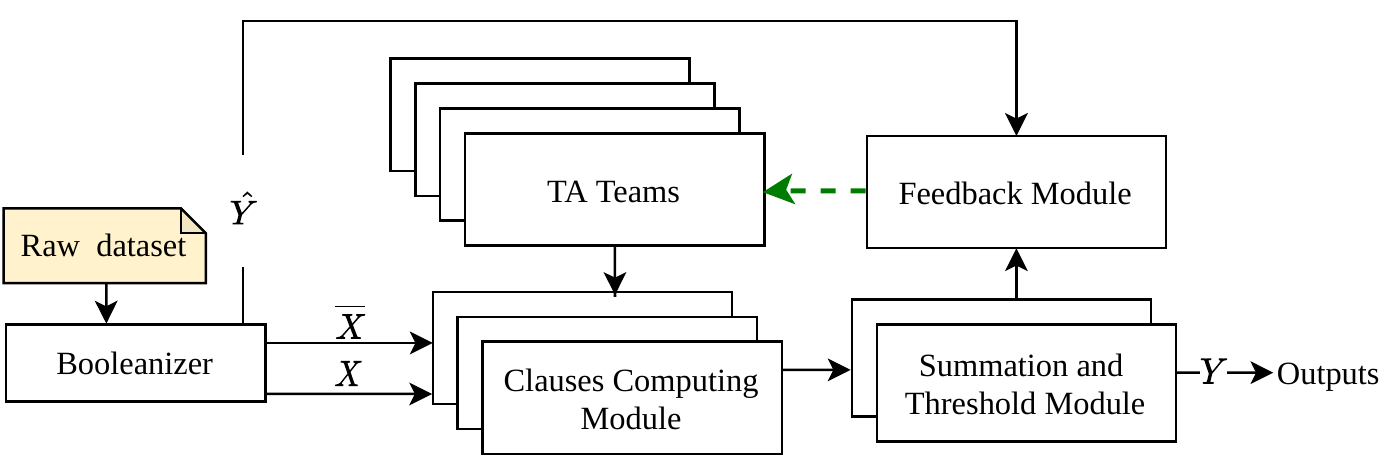}
    \caption{Block diagram of \ac{TM} (dashed green arrow indicates penalties and rewards)\cite{jieleitmnn}}
    \label{fig:tm_block}
\end{figure}

The core components of the Tsetlin Machine are: \emph{a team of Tsetlin Automata (TA) in each clause}, \emph{conjunctive clauses}, \emph{summation and threshold} module and the \emph{feedback} module, as seen in Figure \ref{fig:tm_block}. The TA are \ac{FSM}s that are used to form the propositional logic based relationships that describe an output class through the \textit{inclusion} or \textit{exclusion} of input features and their complements. The states of the TAs for each feature and its compliment are then aligned to a stochastically independent clause computation module. Through a voting mechanism built into the summation and threshold module the expected output class ${Y}$ is generated. During the training phase this class is compared against the target class $\hat{Y}$ and the TA states are incremented or decremented accordingly (this is also referred to as as issuing rewards or penalties).

A fundamental difference between the \ac{TM} and \ac{NN}s is the requirement of a \emph{Booleanizer} module. The key premise is to convert the raw input features and their complements to Boolean features rather than binary encoded features as seen with \ac{NN}s. These Boolean features are also referred to as literals: $\hat{X}$ and ${X}$. Current research has shown that significance-driven Booleanization of features for the Tsetlin Machine is vital in controlling the Tsetlin Machine size and processing requirements~\cite{Wheeldon2020a}. Increasing the number of features will increase the number of TA and increase computations for the clause module and subsequently the energy spent in incrementing and decrementing states in the feedback module. The choice of the number of clauses to represent the problem is also available as a design knob, which also directly affects energy\slash accuracy tradeoffs~\cite{jieleitmnn}. 

\begin{figure}[htbp]
    \centering
    \includegraphics[width=0.3\columnwidth]{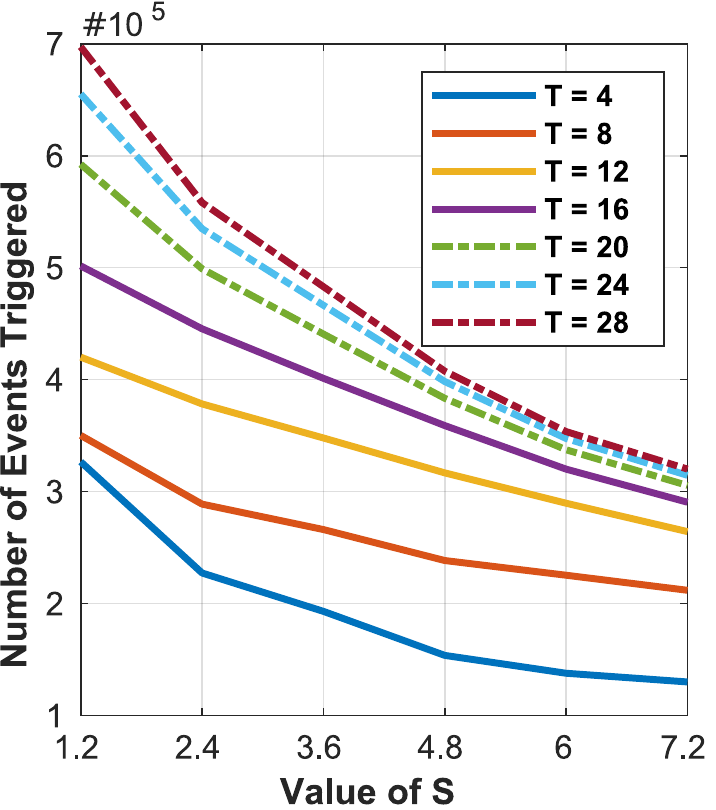}
    \caption{The affect of $T$ and $s$ on reinforcements on \ac{TM}\cite{jieleitmnn}}
    \label{fig:tm_st_stoca}
\end{figure}

The Tsetlin Machine also has two hyper parameters, the \emph{s} value and the \emph{Threshold (T)}. The Threshold parameter is used to determine the clause selection to be used in the voting mechanism, larger Thresholds will mean more clauses partake in the voting and influence the feedback to TA states. The \emph{s} value is used to control the fluidity with which the TAs can transition between states. Careful manipulation of these parameters can be used to determine the flexibility of the feedback module and therefore control the \ac{TM}s learning stability \cite{shafikexplainability}. As seen in Figure \ref{fig:tm_st_stoca}, increasing the Threshold and decreasing the \emph{s} value will lead to more events triggered as more states are transitioned. These parameters must be carefully tuned to balance energy efficiency through minimizing events triggered, and achieving good performance through finding the optimum \emph{s}-\emph{T} range for learning stability in the \ac{KWS} application.    

In order to optimize the TM for \ac{KWS}, due diligence must be given to designing steps that minimize the Boolean feature set. This allows for finding a balance between performance and energy usage through varying the TM hyper parameters and the number of clause computation modules. Through exploitation of these relationships and properties of the \ac{TM}, the \ac{KWS} pipeline can be designed with particular emphasis on feature extraction and minimization of the number of the \ac{TM}s clause computation modules. An extensive algorithmic description of Tsetlin Machine can be found in~\cite{Granmo2018}. The following section will detail how these ideas can be implemented through audio pre-processing and Booleanization techniques for \ac{KWS}.

\section{Audio Pre-processing Techniques for \ac{KWS}}
When dealing with audio data, the fundamental design efforts in pre-processing should be to find the correct balance between reducing data volume and preserving data veracity. That is, while removing the redundancies from the audio stream the data quality and completeness should be preserved. This is interpreted in the proposed \ac{KWS}-\ac{TM} pipeline through two methods: feature extraction through \ac{MFCC}s, followed by discretization control through quantile based binning for Booleanization. These methods are expanded below.    

\subsection{Audio Feature Extraction using \ac{MFCC}}\label{sec:explainMFCC}

\begin{figure}[ht]	
\centering
\includegraphics[width=1\columnwidth]{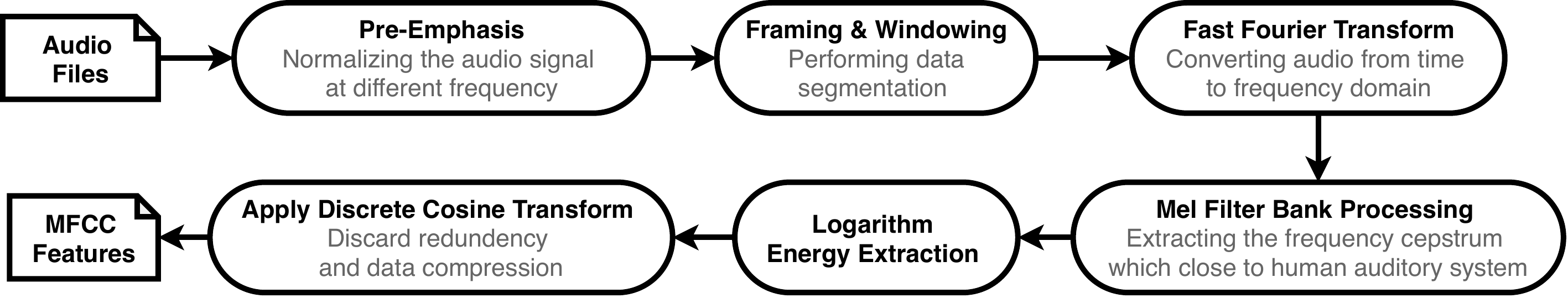}
\caption{\ac{MFCC} pipeline.\label{fig:wsp_mfcc_pipe}}
\end{figure}

Audio data streams are always subject to redundancies in the channel that formalize as nonvocal noise, background noise and silence \cite{5109766, MUSHTAQ2020107389}. Therefore the challenge becomes identification and extraction of the desired linguistic content (the keyword) and maximally discarding everything else. To achieve this we must consider transformation and filtering techniques that can amplify the characteristics of the speech signals against the background information. This is often done through the generation of \ac{MFCC}s as seen in the signal processing flow in Figure \ref{fig:wsp_mfcc_pipe}.

The \ac{MFCC} is a widely used audio file pre-processing method for speech related classification applications \cite{9063000,MUSHTAQ2020107389,DENG202022,8977176,Kaur2015FeatureEA,8936893}. The component blocks in the \ac{MFCC} pipeline are specifically designed for extracting speech data taking into account the intricacies of the human voice. 

The \emph{Pre-Emphasis step} is used to compensate for the structure of the human vocal tract and provide initial noise filtration. When producing glottal sounds when speaking, higher frequencies are damped by the vocal tract which can be characterized as a step roll-off in the signals' frequency spectrum \cite{Picone93}. The Pre-Emphasis step, as its name-sake suggests, amplifies (adds emphasis to) the energy in the high frequency regions, which leads to an overall normalization of the signal \cite{Kamath2019}.     

Speech signals hold a quasi-stationary quality when examined over a very short time period, which is to say that the statistical information it holds remains near constant \cite{5109766}. This property is exploited through the \emph{Framing and Windowing} step. The signal is divided into around 20ms frames, then around 10-15ms long window functions are multiplied to these overlapping frames, in doing so we preserve the temporal changes of the signal between frames and minimize discontinuities (this is realized through the smoothed spectral edges and enhanced harmonics of the signal after the subsequent transformation to the frequency domain) \cite{10.1002/9781118142882.part6}. The windowed signals are then transformed to the frequency domain through a Discrete Fourier Transform (DFT) process using the \emph{Fast Fourier Transform (FFT)} algorithm. FFT is chosen as it is able to find the redundancies in the DFT and reduce the amount of computations required offering quicker run-times. 

The human hearing system interprets frequencies linearly up to a certain range (around 1KHz) and logarithmically thereafter. Therefore, adjustments are required to translate the FFT frequencies to this non-linear function \cite{NALINI20161}. This is done through passing signal through the \emph{Mel Filter Banks} in order to transform it to the \emph{Mel Spectrum} \cite{9031431}. The filter is realized by overlapping band-pass filters to create the required warped axis. Next, the logarithm of the signal is taken, this brings the data values closer and less sensitive to the slight variations in the input signal \cite{9031431}. Finally we perform a \emph{Discrete Cosine Transform (DCT)} to take the resultant signal to the \emph{Cepstrum} domain \cite{8721379}. The DCT function is used as energies present in the signal are very correlated as a result of the overlapping Mel filterbanks and the smoothness of the human vocal tract; the DCT finds the co-variance of the energies and is used to calculate the \ac{MFCC} features vector \cite{Kamath2019,JOTHILAKSHMI20099799}. This vector can be passed to the Booleanizer module to produce the input Boolean features, as described next.

\subsection{Feature Booleanization}\label{sec:explainBoolMethod}

As described in Section~\ref{sec:wsp_design_tm}, Booleanization is an essential step for logic based feature extraction in Tsetlin Machines. Minimizing the Boolean feature space is crucial to the Tsetlin Machine's optimization. The size and processing volume of a TM is primarily dictated by the number of Booleans \cite{Wheeldon2020a}. Therefore, a pre-processing stage for the audio features must be embedded into the pipeline before the \ac{TM} to allow for granularity control of the raw \ac{MFCC} data. The number of the Booleanized features should be kept as low as possible while capturing the critical features for classification \cite{Wheeldon2020a}.

The discretization method should be able to adapt to, and preserve the statistical distribution of the \ac{MFCC} data. The most frequently used method in categorizing data is through binning. This is the process of dividing data into group, individual data-points are then represented by the group they belong to. Data points that are close to each other are put into the same group thereby reducing data granularity ~\cite{Granmo2018}. Fixed width binning methods are not effective in representing skewed distribution and often result in empty bins, they also require manual decision making for bin boundaries. 

Therefore, for adaptive and scalable Booleanization quantile based binning is preferred. Through binning the data using its own distribution, we maintain its statistical properties and do not need to provide bin boundaries, merely the number of bins the data should be discretized into. The control over the number of quantiles is an important parameter in obtaining the final Boolean feature set. Choosing two quantiles will result in each MFCC coefficient being represented using only one bit; however, choosing ten quantiles (or bins) will result in four bits per coefficient. Given the large number of coefficients present in the \ac{KWS} problem, controlling the number of quantiles is an effective way to reduce the total \ac{TM} size.

\subsection{The \ac{KWS}-\ac{TM} pipeline}\label{sec:explainMFCCtmPipe}
\begin{figure}[ht]	
\includegraphics[width=\columnwidth]{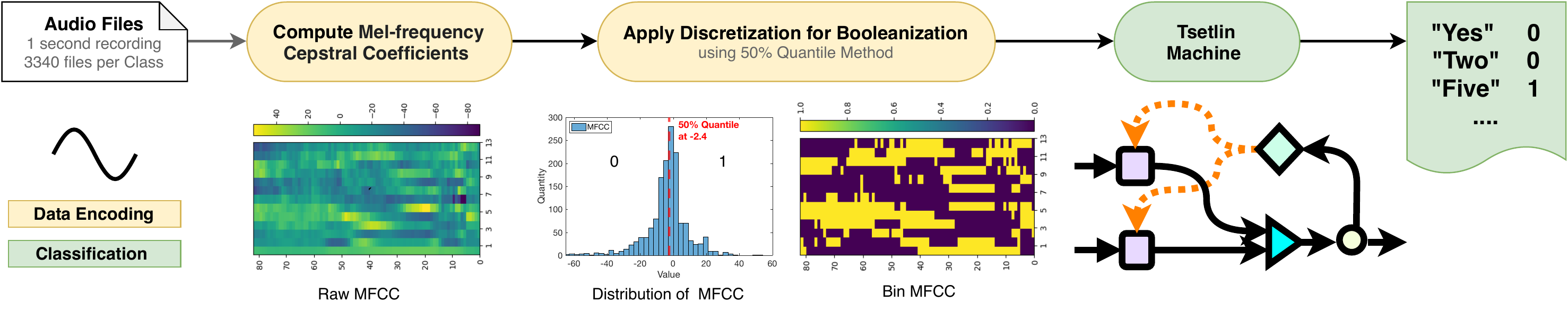}
\caption{The data encoding and classification stages in the \ac{KWS}-\ac{TM} pipeline}
\label{fig:wsp_mfcc_tm}
\end{figure}

The \ac{KWS}-\ac{TM} pipeline is composed of the the data encoding and classification blocks presented in Figure \ref{fig:wsp_mfcc_tm}. The data encoding scheme encompasses the generation of \ac{MFCC}s and the quantile binning based Booleanization method. The resulting Booleans are then fed to the Tsetlin Machine for classification. The figure highlights the core attributes of the pre-processing blocks: the ability to extract the audio features only associated with speech through \ac{MFCC}s and the ability to control their Boolean granularity through quantile binning.

To explore the functionality of the pipeline and the optimizations that can be made, we return to our primary intentions, i.e., to achieve energy efficiency and high learning efficacy in \ac{KWS} applications. We can now use the design knobs offered in the pre-processing blocks, such as variable window size in the MFCC generation, and control over the number of quantiles to understand how these parameters can be used in presenting the Boolean data to the \ac{TM} in a way to returns good performance utilizing the least number of Booleans. Through \cref{sec:wsp_design_tm} we have also seen the design knobs available through variation of the hyperparameters \emph{s} and Threshold \emph{T}, as well as the number of clause computation modules used to represent the problem. Varying the parameters in both the encoding and classification stages through an experimental context will uncover the impact they have on the overall \ac{KWS} performance and energy usage.

\section{Experimental Results}
To evaluate the proposed \ac{KWS}-\ac{TM} pipeline, Tensorflow speech command dataset was used\footnote{Tensorflow speech command: https://tinyurl.com/TFSCDS}. The dataset consists of many spoken keywords collected from a variety of speakers with different accents, as well as male and female gender. The datapoints are stored as 1 second long audio files where the background noise is negligible. This reduces the effect of added redundancies in the \ac{MFCC} generation, given our main aim is predominantly to test functionality, we will explore the impact of noisy channels in our future work. This dataset is commonly used in testing the functionality of ML models and will therefore allow for fair comparisons to be drawn \cite{warden2018speech}.     

From the Tensorflow dataset, 10 keywords: "Yes", "No", "Stop", "Seven", "Zero", "Nine", "Five", "One", "Go" and "Two", have been chosen to explore the functionality of the pipeline using some basic command words. Considering other works comparing \ac{NN} based pipelines, 10 keywords is the maximum used \cite{Zhang2018, 8502309}. Among the keywords chosen, there is an acoustic similarity between "No" and "Go", therefore, we explore the impact of 9  keywords together (without "Go") and then the effect of "No" and "Go" together. The approximate ratio of training data, testing data and validation data is given as 8:1:1 respectively with a total of 3340 datapoints per class. Using this setup, we will conduct a series of experiments to examine the impact of the various parameters of the \ac{KWS}-\ac{TM} pipeline discussed earlier. The experiments are as follows:
\begin{itemize}
        \item Manipulating the window length and window steps to control the number of \acp{MFCC} generated. 
        \item Exploring the effect of different quantile bins to change the number of Boolean features.
        \item Using a different number of the keywords ranging from the 2 to 9 to explore the scalability of the pipeline.
        \item Testing the effect on performance of acoustically different and similar keywords.
        \item Changing the size of the \ac{TM} through manipulating the number of clause computation modules, optimizing performance through tuning the  feedback control parameters \emph{s} and \emph{T}. 
\end{itemize}
    
\subsection{\ac{MFCC} Setup} \label{sweepMFCC}\label{sec:expMFCCwin}
It is well defined that the number of input features to the \ac{TM} is one of the major factors that affect its resource usage \cite{shafikexplainability,Wheeldon2020a,jieleitmnn}. Increased raw input features means more Booleans are required to represent them and thus the number of \ac{TA} in the \ac{TM} will also increase leading to more energy required to provide feedback to them. Therefore, reducing the number of features at the earliest stage of the data encoding stage of the pipeline is crucial to implementing energy-frugal \ac{TM} applications. 
    
\begin{figure}[ht]
    \centering
    \begin{tabular}{c}
     \includegraphics[width=0.6\columnwidth]{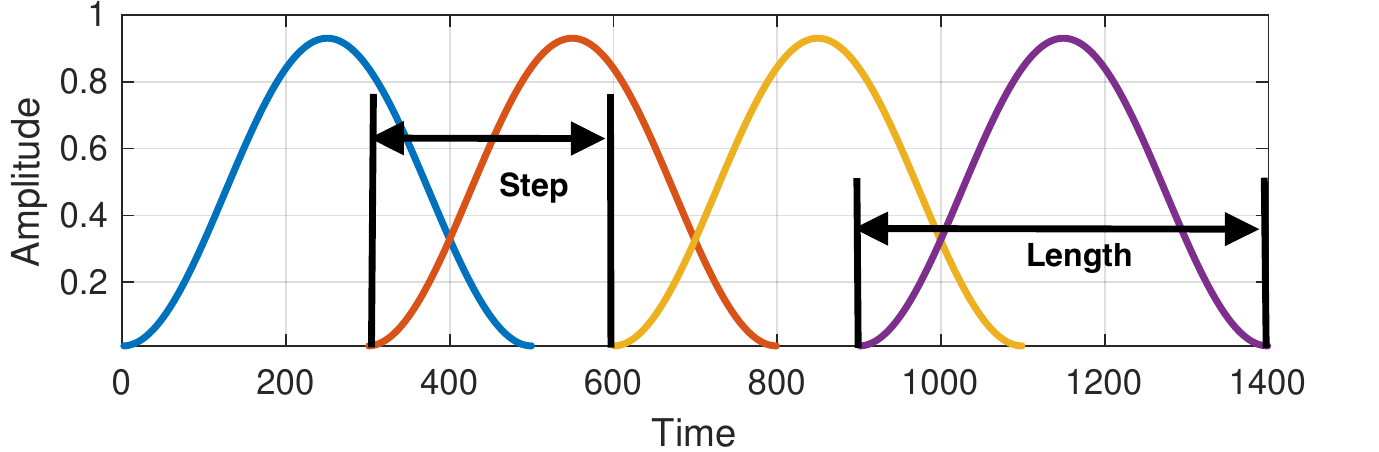} \\ 
     \small{(a) The windowing process.} \\
     \includegraphics[width=0.6\columnwidth]{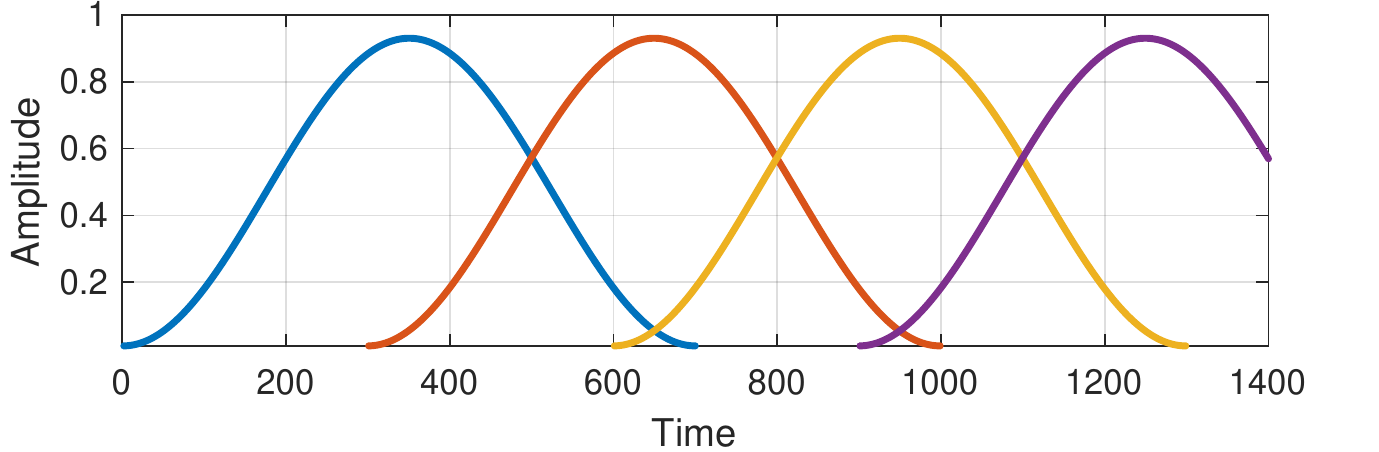} \\ 
     \small{(b) Effect of increasing window length.} \\
     \includegraphics[width=0.6\columnwidth]{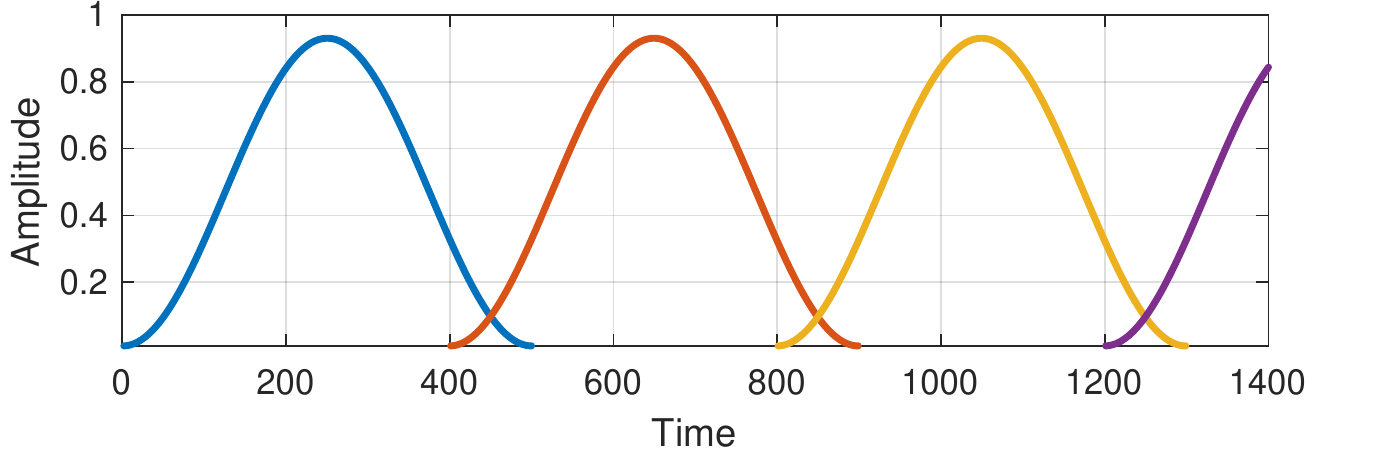}\\
     \small{(c) Effect of increasing window step.}\\
    \end{tabular}
    \caption{The Hamming window function applied to audio pre-processing.} 
    \label{fig:wsp_hamWin_all}
\end{figure}

The first set of parameters available in manipulating the number of features comes in the form of the \emph{Window Step} and the \emph{Window Length} (this takes place in the "Framing an Windowing" stage in \cref{fig:wsp_mfcc_tm}) in \ac{MFCC} generation and can be seen through Figure \ref{fig:wsp_hamWin_all}(a). 

The window function is effective in reducing the spectral distortion by tapering the sample signal at the beginning and ending of each frame (We use overlapping frames to ensure signal continuity is not lost). Smaller Window Steps lead to a more fine grained and descriptive representation of the audio features through more frames and therefore more \acp{MFCC} but this also increases computations and latency. 
    
\begin{figure}[ht] 
    \begin{tabular}{ c c }
    \centering
          \includegraphics[width=0.45\columnwidth]{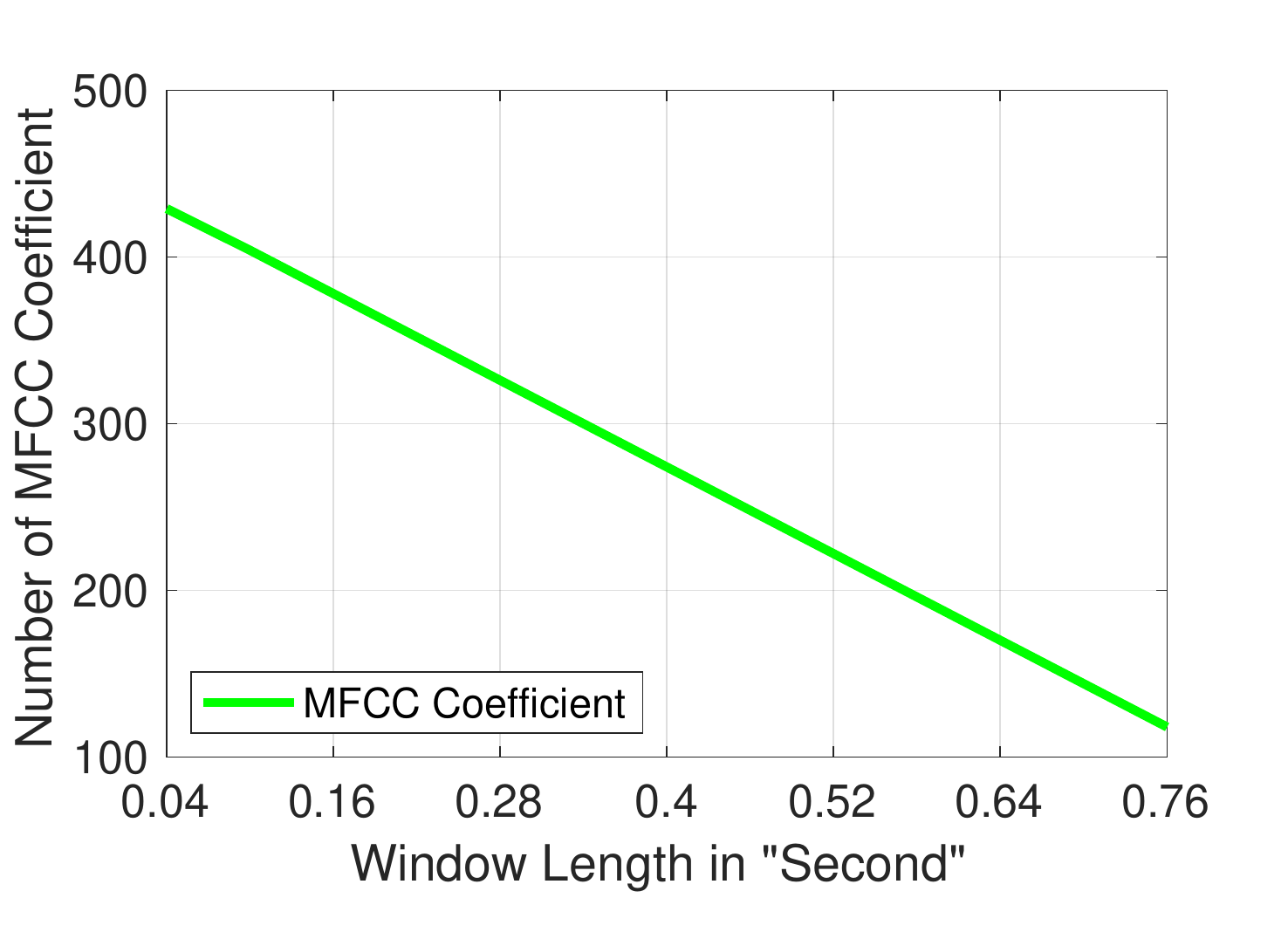} &     \includegraphics[width=0.45\columnwidth]{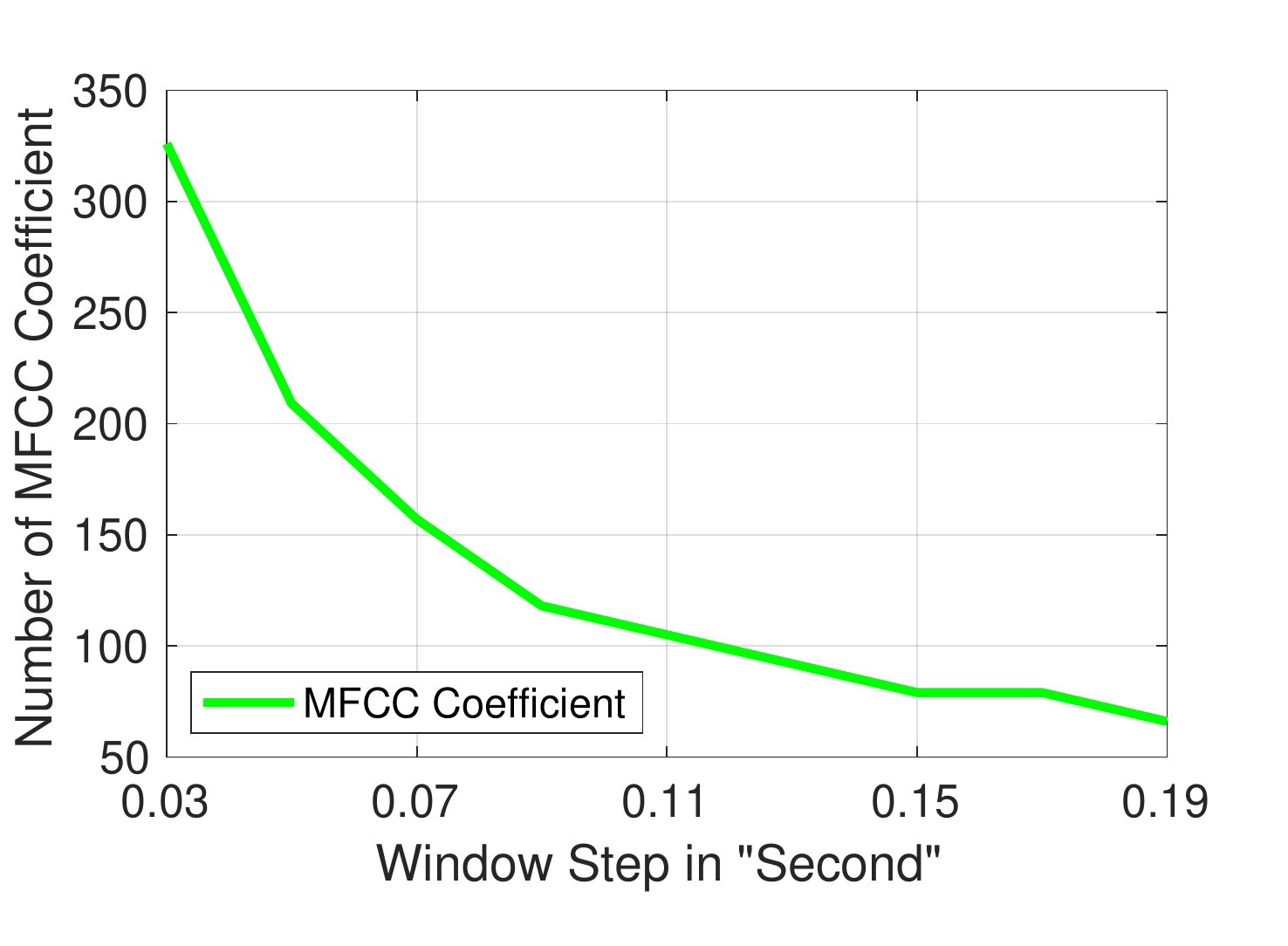}\\
        \small{(a) The effect of increasing window length.} &
         \small{(b) The effect of increasing window step.} 
        \end{tabular}
       \caption{Changing the number of \ac{MFCC} coefficients through manipulating the Window parameters.}
        \label{fig:wsp_win_ALL_numcoe}
\end{figure}

Increasing the Window Length leads to a linear decrease in the total number of frames and therefore the \acp{MFCC} as seen in Figure \ref{fig:wsp_win_ALL_numcoe}(a). Given that the Window Steps are kept constant for this experiment, we have a linearly decreasing number of window overlaps resulting in a linearly decreasing total number of window functions, FFTs and subsequent computations. This leads to the linear decrease in the \acp{MFCC} across all frames.  
    
Increasing the Window Step leads to much sharper fall given the overlapping regions now no longer decrease linearly as seen in Figure \ref{fig:wsp_win_ALL_numcoe}(b). This results in a total number of non-linearly decreasing window functions and therefore much fewer FFTs and so on, leading to much fewer \acp{MFCC} across all frames. As a result of this, the smaller the increase in the Window Step the larger the decrease in the number of frames and therefore \acp{MFCC}.   
    
\begin{figure}[ht] 
    \begin{tabular}{ c c }
    \centering
         \includegraphics[width=0.45\columnwidth]{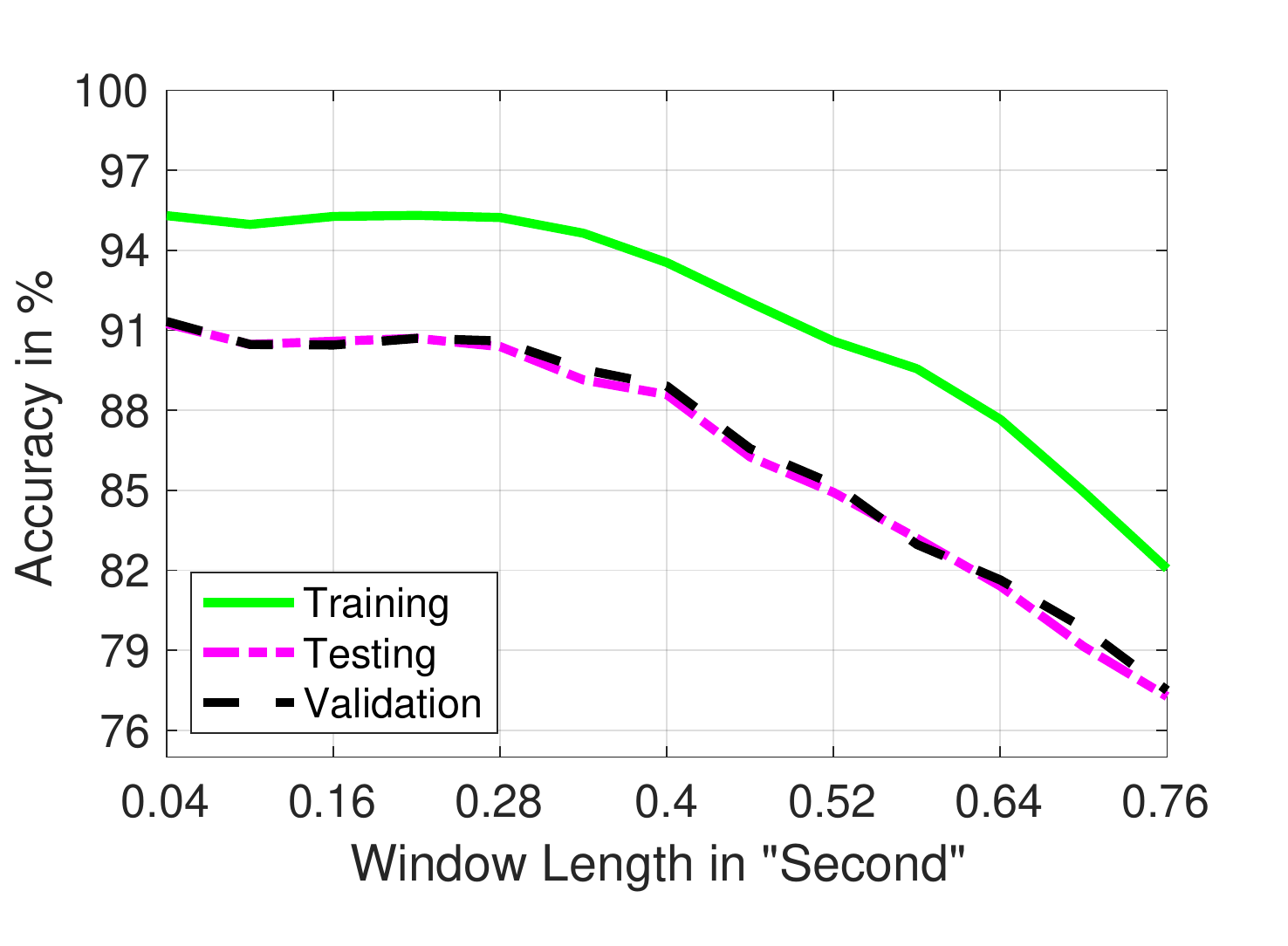} &     \includegraphics[width=0.45\columnwidth]{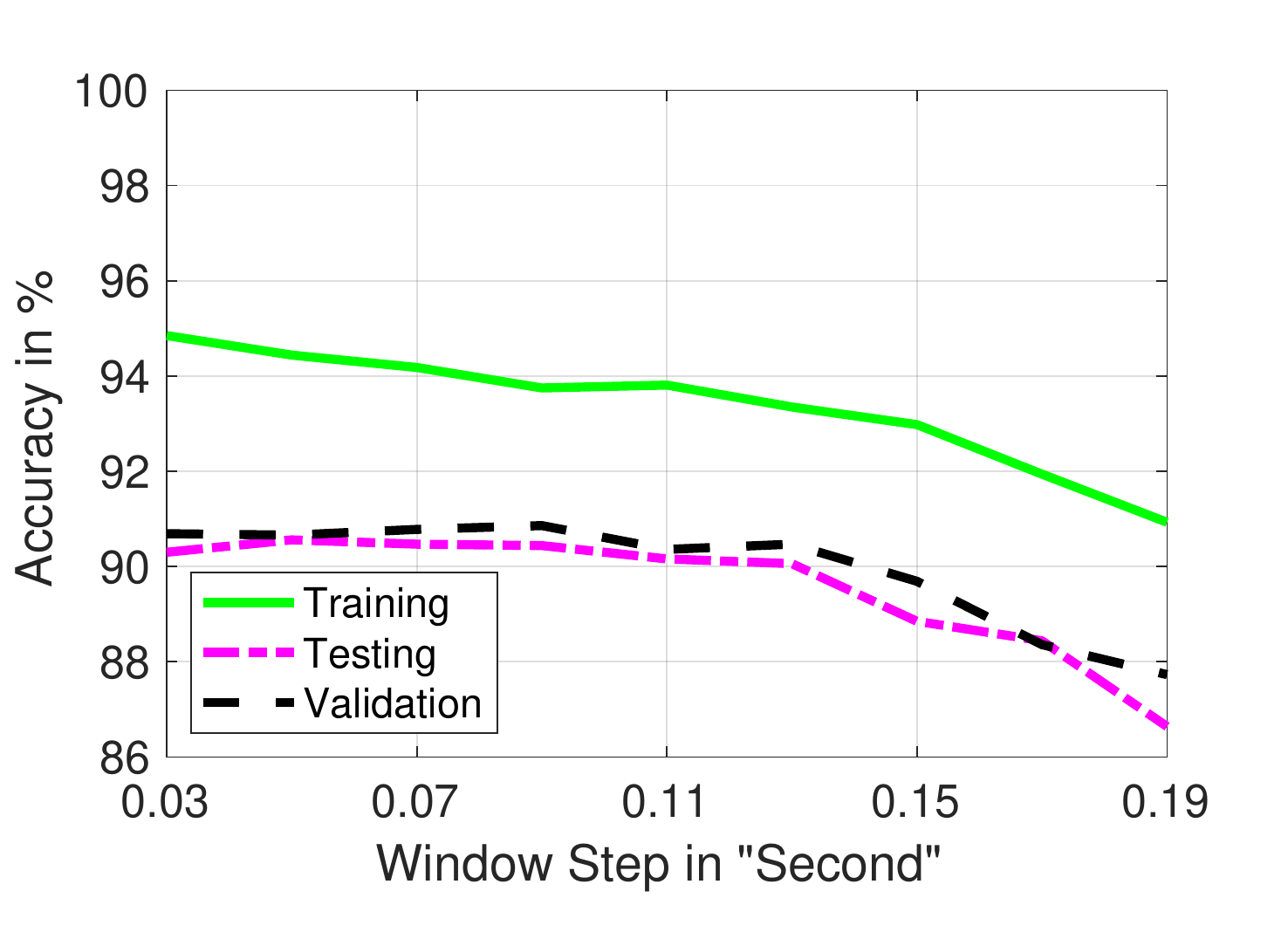}\\
         \small{(a) Effect of window length on accuracy.} &
         \small{(b) Effect of window step on accuracy.}
        \end{tabular}
        \caption{Effect of changing window parameters on classification accuracy.}
        \label{fig:wsp_win_ALL_accu}
\end{figure}

To test the effectiveness of manipulating the Window Length and Window Step, the MFCC coefficients were produced for 4 keywords and the \ac{TM}s classification performance was examined as seen in Figure \ref{fig:wsp_win_ALL_accu}(a) and Figure \ref{fig:wsp_win_ALL_accu}(b). Changing the Window Length results in much bigger falls in accuracy compared to Window Step. This is due to the diminished signal amplitudes at the window edges, longer windows mean more tapering of the edge amplitudes and fewer overlaps to preserve the signal continuities as seen through Figure \ref{fig:wsp_hamWin_all}(b). As a result, the fidelity of generated the \ac{MFCC} features is reduced.

The effect of increasing the Window Step leads to a smaller drop in accuracy.  We see the testing and validation accuracy remain roughly the same at around 90.5$\%$ between 0.03 and 0.10 second Window Steps and then experience a slight drop. Once again this is due to the tapering effect of the window function, given the window length remains the same for this experiment, we know that the increasing of window steps will mean far fewer total overlaps and a shrinking overlapping region as seen in Figure \ref{fig:wsp_hamWin_all}(c). The overlaps are used to preserve the continuity of the signal against the window function edge tapering, as the size of the overlapping regions decrease, the effect of edge tapering increases thereby leading to increased loss of information. The accuracy remains constant up to a Window Step of 0.1s as the Window Length is sufficiently long to capture enough of the signal information, once the overlapping regions start to shrink we experience the loss in accuracy.    

We can see that increasing the Window Step is very effective in reducing the number of frames and therefore the total number \ac{MFCC} coefficients across all frames and providing the Window Length is long enough, the reduction in performance is minimal. To translate these findings toward energy efficient implementations, we must give increased design focus to finding the right balance between the size of the Window Step parameter and the achieved accuracy given the reduction in computations from the reduction in features produced.

\subsection{Impact of Number of Quantiles}\label{sec:expBoolMethod}

\begin{figure}[ht]	
    \includegraphics[width=1\columnwidth]{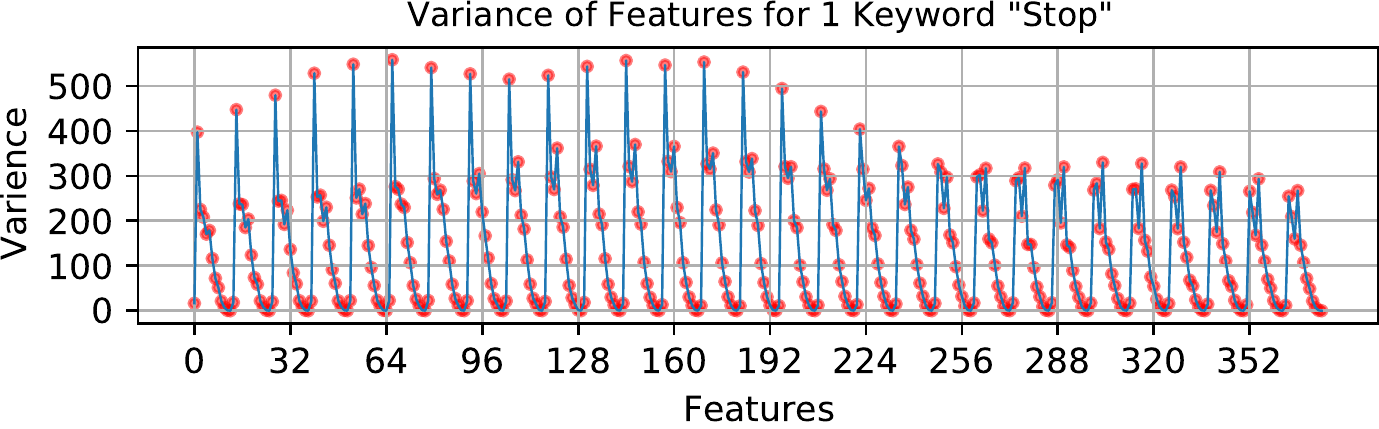}
    \caption{Variance between \ac{MFCC} features.\label{fig:wsp_varience1keySTOP_line}}
\end{figure}

\begin{figure}[ht]	
    \includegraphics[width=1\columnwidth]{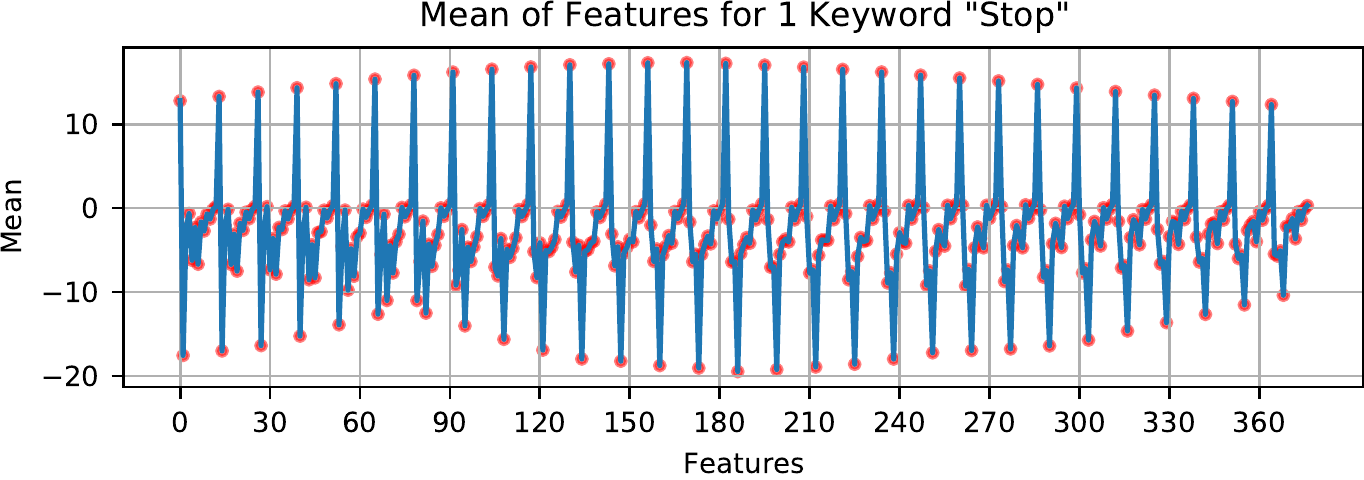}
    \caption{Mean of MFCC features.\label{fig:wsp_Mean1keySTOP_scatter}}
\end{figure}

Increased granularity through more bins will lead to improved performance but it is observed that this is not the case entirely. Table \ref{tab:wsp_diffbool} shows the impact of the \ac{KWS}-\ac{TM} performance when increasing the number of bins. The testing and validation accuracy remain around the same with 1 Boolean per feature compared with 4 Booleans per feature. Figure \ref{fig:wsp_varience1keySTOP_line} shows the large variance in some feature columns and no variance in others. The zero variance features are redundant in the subsequent Booleanization, they will be represented through the same Boolean sequence. The features with large variances are of main interest. We see that the mean for these features is relatively close to zero compared to their variance (as seen in Figure \ref{fig:wsp_Mean1keySTOP_scatter}), therefore one Boolean per feature representation is sufficient, a 1 will represent values above the mean and 0 will represent below. The logical conclusion to be made from these explorations is that the \ac{MFCC} alone is sufficient in both eliminating redundancies and extracting the keyword properties and does not require additional granularity beyond one Boolean per feature to distinguish classes.

    We have seen that the large variance of the \acp{MFCC} mean that they are easily represented by 1 Boolean per feature and that is sufficient to achieve high performance. This is an important initial result, for offline learning we can now also evaluate the effect of removing the no variance features in future work to further reduce the total number of Booleans. From the perspective of the Tsetlin Machine there is an additional explanation as to why the performance remains high even when additional Boolean granularity is allocated to the \ac{MFCC} features. Given that there are a large number datapoints in each class (3340), if the \ac{MFCC}s that describe these datapoints are very similar then the \ac{TM} will have more than sufficient training data to settle on the best propositional logic descriptors. This is further seen by the high training accuracy compared to the testing and validation accuracy.  

    \vspace{1em}
    \begin{table}[htbp]
        \caption{Impact of increasing quantiles with 4 classes}
        \centering
        \begin{tabular}{c|c|c|c|c|c}
        Training & Testing & Validation & Num.Bins & Bools per Feature & Total Bools\\
        \hline
        \hline
                  94.8\%  & 91.3\%  & 91.0\%   &  2   & 1 & 378\\
                  96.0\%  & 92.0\%  & 90.7\%   &  4   & 2 & 758\\
                  95.9\%  & 90.5\%  & 91.0\%   &  6   & 3 & 1132\\
                  95.6\%  & 91.8\%  & 92.0\%   &  8   & 3 & 1132\\
                  97.1\%  & 91.0\%  & 90.8\%   &  10  & 4 & 1512\\
        \hline
        \end{tabular}
                \newline
        \label{tab:wsp_diffbool}
    \end{table}

\subsection{Impact of Increasing the Number of Keywords}\label{sec:expNumKeys}
\begin{figure}[ht] 
    \begin{tabular}{ c c }
    \centering
         \includegraphics[width=0.45\columnwidth]{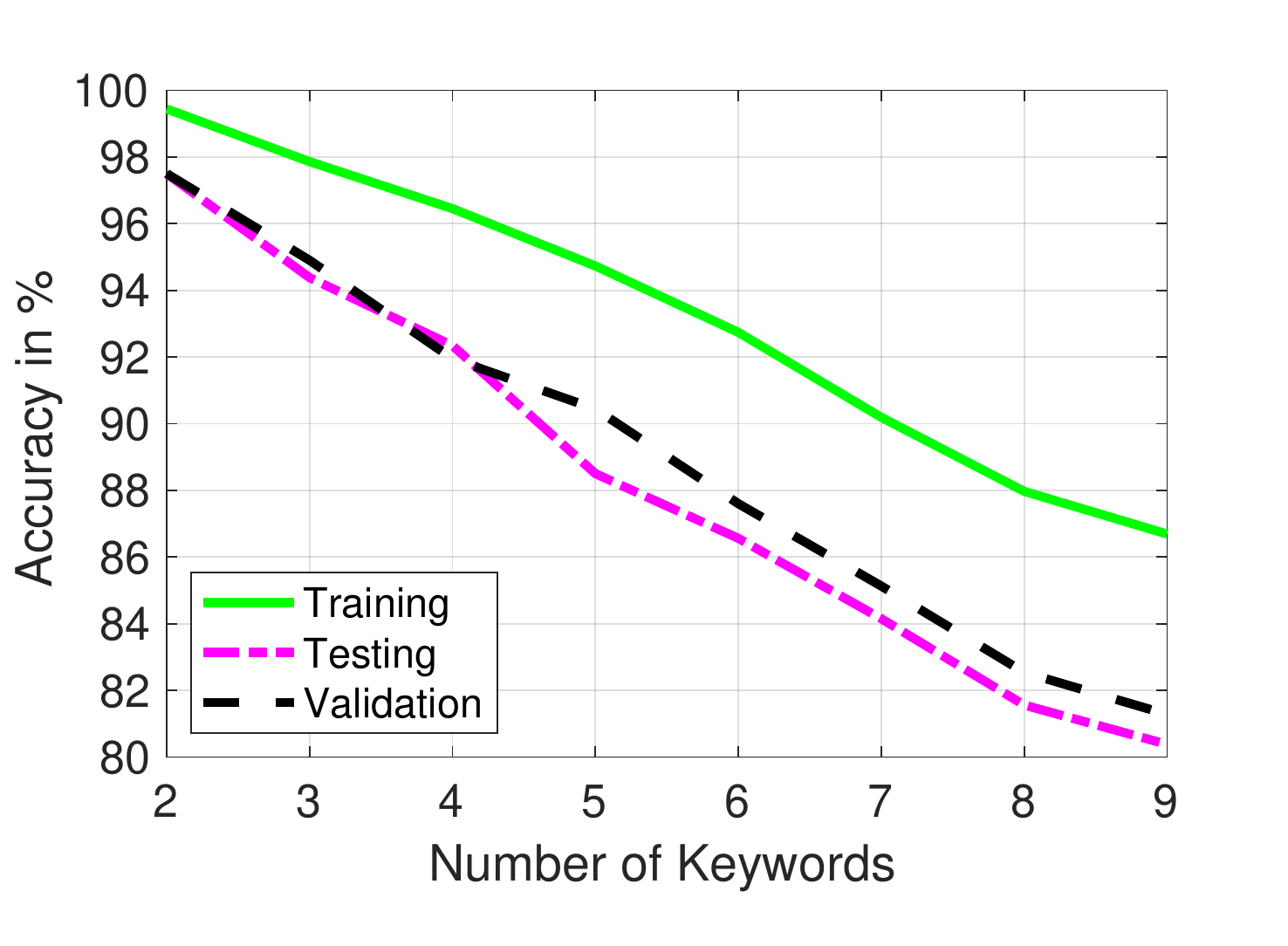} &     \includegraphics[width=0.45\columnwidth]{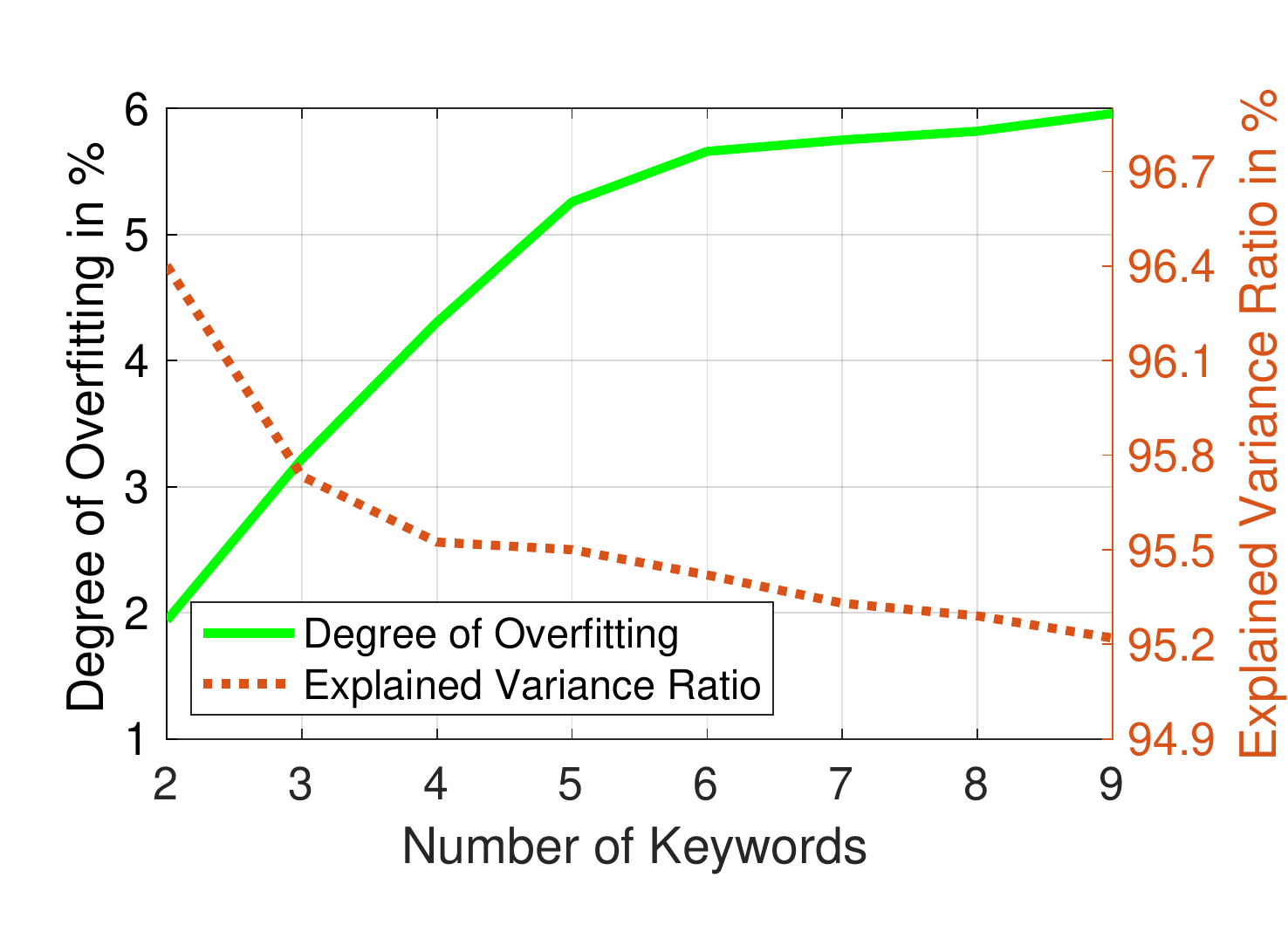}\\
         \small{(a) The effect on accuracy.} &
         \small{(b) The amount of overfitting.}
        \end{tabular}
        \caption{The effect of increasing the number of keywords.}
        \label{fig:wsp_numKey_Accu_N_OverF}
    \end{figure}

\cref{fig:wsp_numKey_Accu_N_OverF}(a) shows the linear nature with which the training, testing and validation accuracy decrease as the number of keywords are increased for a \ac{TM} with 450 clauses with 200 epochs for training. We note that the testing and validation accuracy start to veer further away from the training accuracy with the increase of keywords. This performance drop is expected in \ac{ML} methods as the problem scales \cite{ESCcnn19-1}. Despite the large number of datapoints per keyword this is an indicator of overfitting, as confirmed through \cref{fig:wsp_numKey_Accu_N_OverF}(b) showing around a 4$\%$ increase. The implication of this is that increased number of keywords make it difficult for the \ac{TM} to create distinct enough propositional logic to separate the classes. The performance drop is caused when the correlation of keywords outweighs the number of datapoints to distinguish each of them. This behavior is commonly observed in \ac{ML} models for audio classification applications \cite{DENG202022}.

\begin{figure}[ht]	
    \centering
    \includegraphics[width=7 cm]{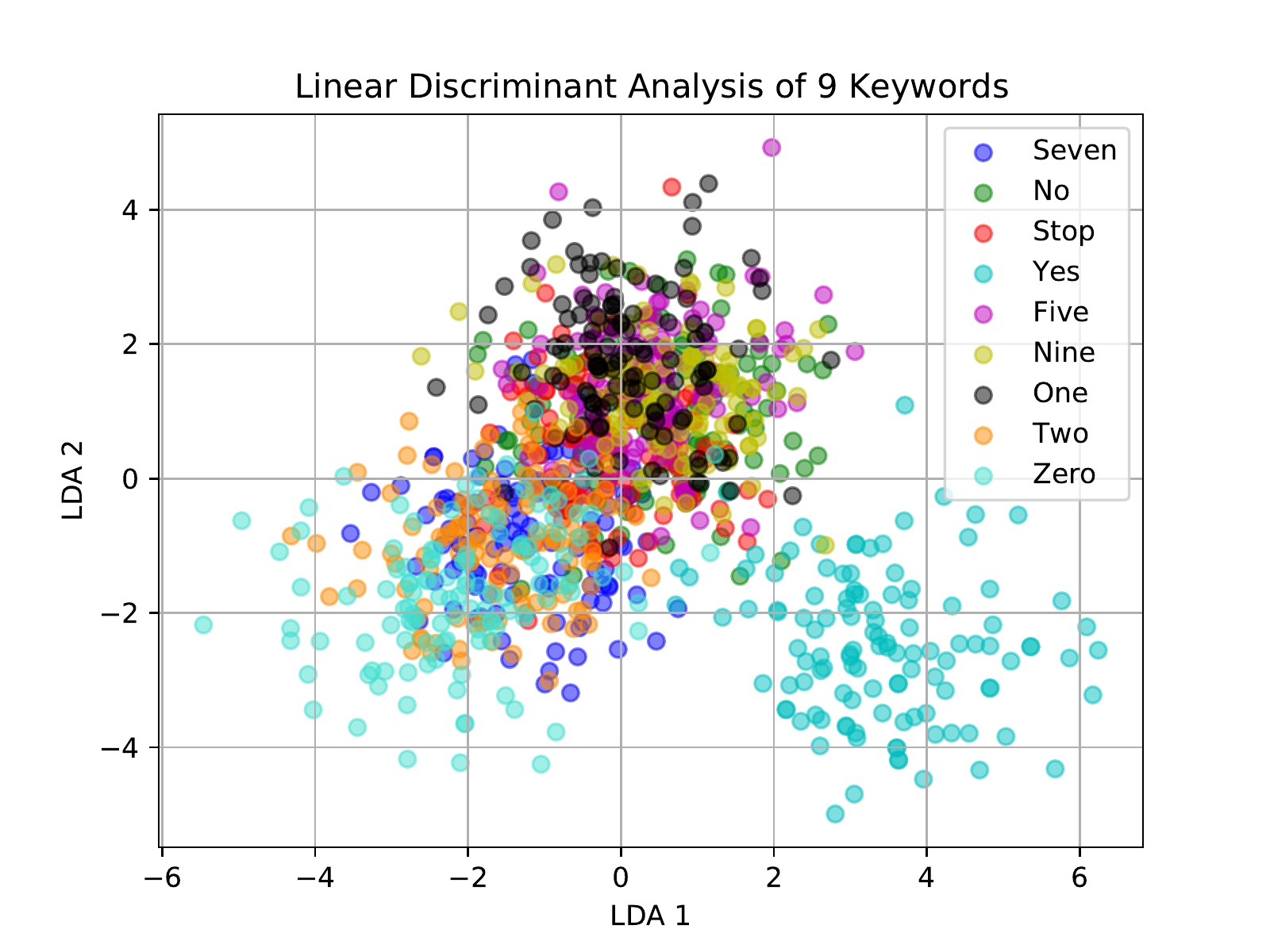}
    \caption{LDA of 9 keywords.\label{fig:wsp_lda_all9keys}}
\end{figure}

The explained variance ratio of the dataset with an increasing number of keywords was taken for the first 100 Principle Component Analysis eigenvalues, as seen in \cref{fig:wsp_numKey_Accu_N_OverF}(b). We observe that as the number of keywords is increased, the system variance decreases, i.e. the inter-class features start to become increasingly correlated. Correlated inter-class features will lead to class overlap and degrade \ac{TM} performance \cite{Wheeldon2020a}. Through examination of the two largest Linear Discriminant component values for the 9 keyword dataset, we clearly see in Figure \ref{fig:wsp_lda_all9keys} that there is very little class separability present.          

To mitigate against the effect on performance of increasing keywords, there are two methods available: Firstly to adjust the Tsetlin Machines hyperparameters to enable more events triggered (see Figure \ref{fig:tm_st_stoca}). In doing so the this may allow the \ac{TM} to create more differing logic to describe the classes. Then, by increasing the number of clause computation modules, the \ac{TM} will have a larger voting group in the Summation and Threshold module and potential reach the correct classification more often. Secondly the quantity of the datapoints can be increased, however, for this to be effective the new dataset should hold more variance and completeness when describing each class. This method of data regularization is often used in audio \ac{ML} applications to deliberately introduce small variance between datapoints \cite{MUSHTAQ2020107389}.

\subsection{Acoustically Similar Keywords}\label{sec:expKeysSoundsLike}

In order to test the robustness of the \ac{KWS}-\ac{TM} pipeline functionality, we must emulate real-word conditions where a user will use commands that are acoustically similar to others. \cref{tab:wsp_tab_simikey} shows the results of such circumstances. The \emph{Baseline} experiment is a \ac{KWS} dataset consisting of 3 keywords: 'Yes', 'No' and 'Stop'. The second experiment then introduces the keyword 'Seven' to the dataset and the third experiment introduces the keyword 'Go'.
    
\begin{figure}[ht] 
    \begin{tabular}{ c c }
    \centering
         \includegraphics[width=0.45\columnwidth]{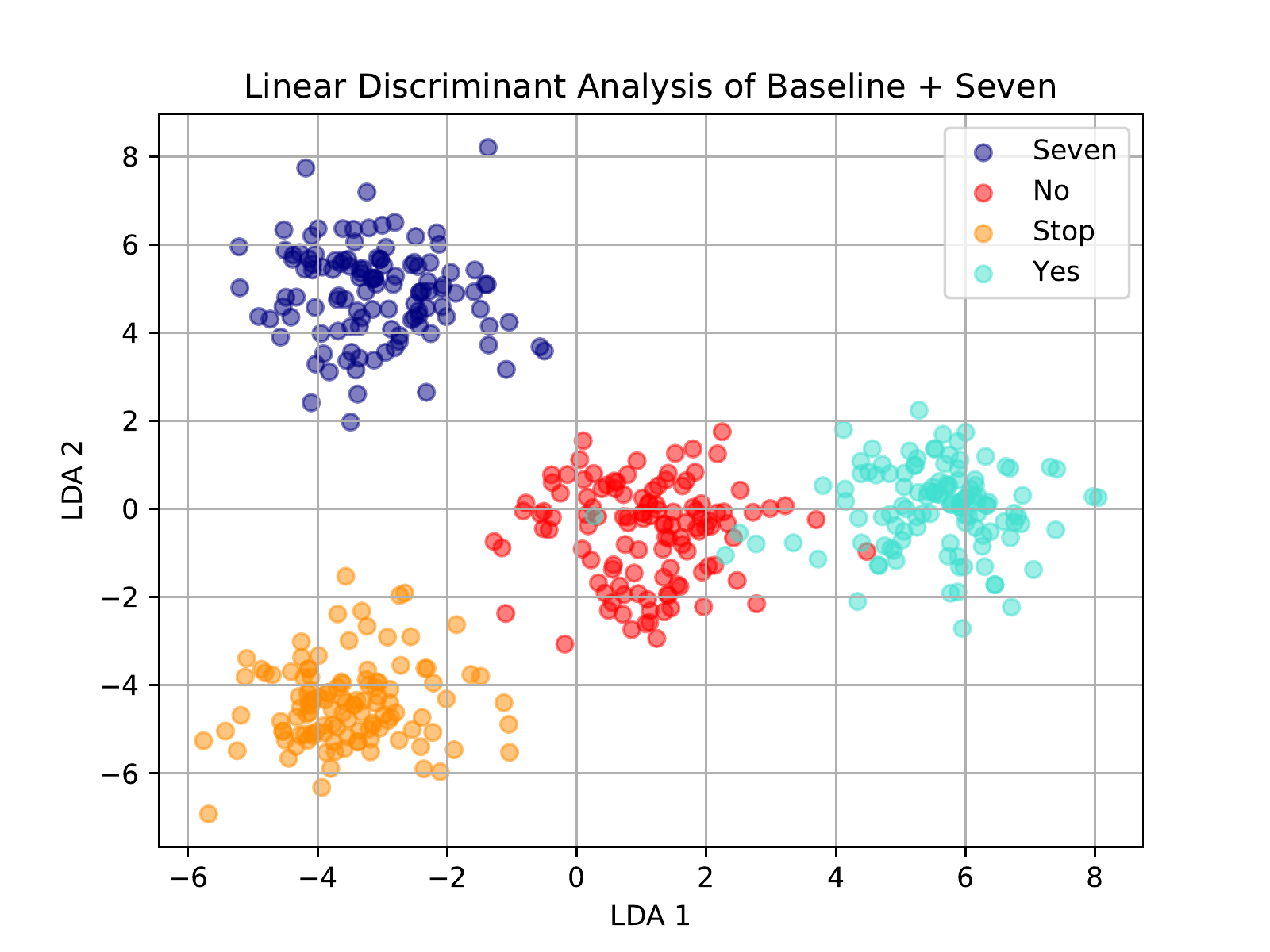} &     \includegraphics[width=0.45\columnwidth]{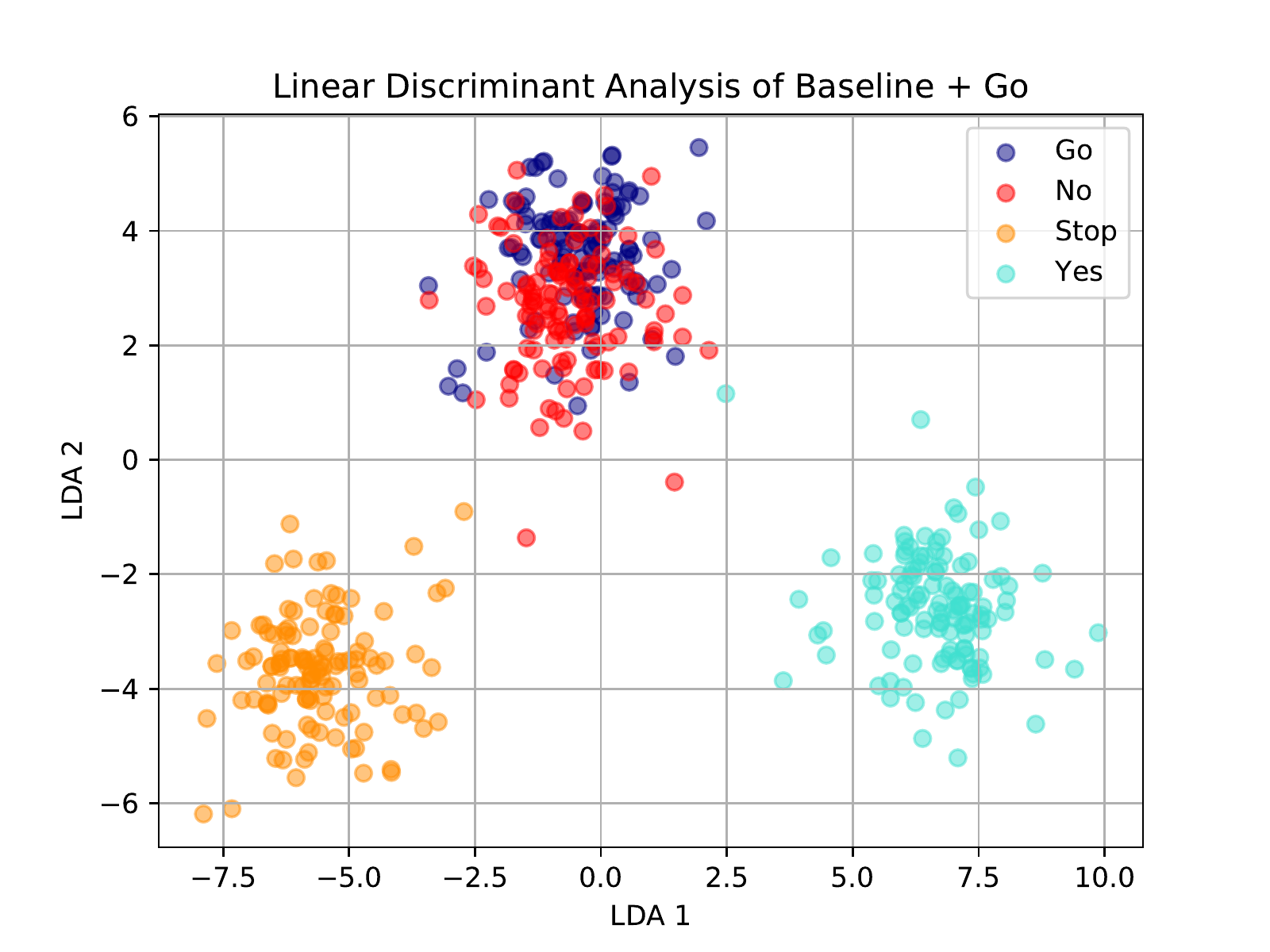}\\
         \small{(a) The Baseline with `Seven'.} &
         \small{(b) The Baseline with `Go'.} 
        \end{tabular}
        \caption{The LDA of 4 keywords - the Baseline with one other.}
        \label{fig:wsp_lda_SevnGo_4keys_all}
    \end{figure}

The addition of 'Seven' causes a slight drop in accuracy adhering to our previously made arguments of increased correlation and the presence of overfitting. However the key result is the inclusion of 'Go'; 'Go' is acoustically similar to 'No' and this increases the difficulty in separating these two classes. We see from Figure \ref{fig:wsp_lda_SevnGo_4keys_all}(a), showing the first two LDA components that adding 'Seven' does not lead to as much class overlap as adding 'Go' as seen in Figure \ref{fig:wsp_lda_SevnGo_4keys_all}(b). As expected, the acoustic similarities of 'No' and 'Go' lead to significant overlap. We have seen from the previous result (Figure \ref{fig:wsp_lda_all9keys}) that distinguishing class separability is increasingly difficult when class overlaps are present.

\begin{table}[htbp]
        \centering

        \caption{Impact of acoustically similar keywords.}
        \begin{tabular}{l||c|c|c}
        Experiments  &Training & Testing & Validation \\
        \hline
        \hline
        Baseline            & 94.7\%  & 92.6\%  & 93.1\%  \\
        Baseline + `Seven'  & 92.5\%  & 90.1\%  & 90.2\%  \\
        Baseline + `Go'     & 85.6\%  & 82.6\%  & 80.9\%   \\
                \hline
        \end{tabular}
                \newline
        \label{tab:wsp_tab_simikey}
\end{table}

    \subsection{Number of Clauses per Class} \label{sec:sweepClause}
    So far we have considered the impact of Booleanization granularity, problem scalabilty and robustness when dealing with acoustically similar classes. Now, we turn our attention towards optimizing the  \ac{KWS}-\ac{TM} pipeline to find the right functional balance between performance and energy efficiency. This is made possible through two streams of experimentation: manipulating the number of clauses for each keyword class in the \ac{TM} and observing the energy expenditure and accuracy, and experimenting with the \ac{TM}s hyperparameters to enable better performance using fewer clauses.   
    
    \begin{figure}[ht] 
    \begin{tabular}{ c c }
    \centering
         \includegraphics[width=0.45\columnwidth]{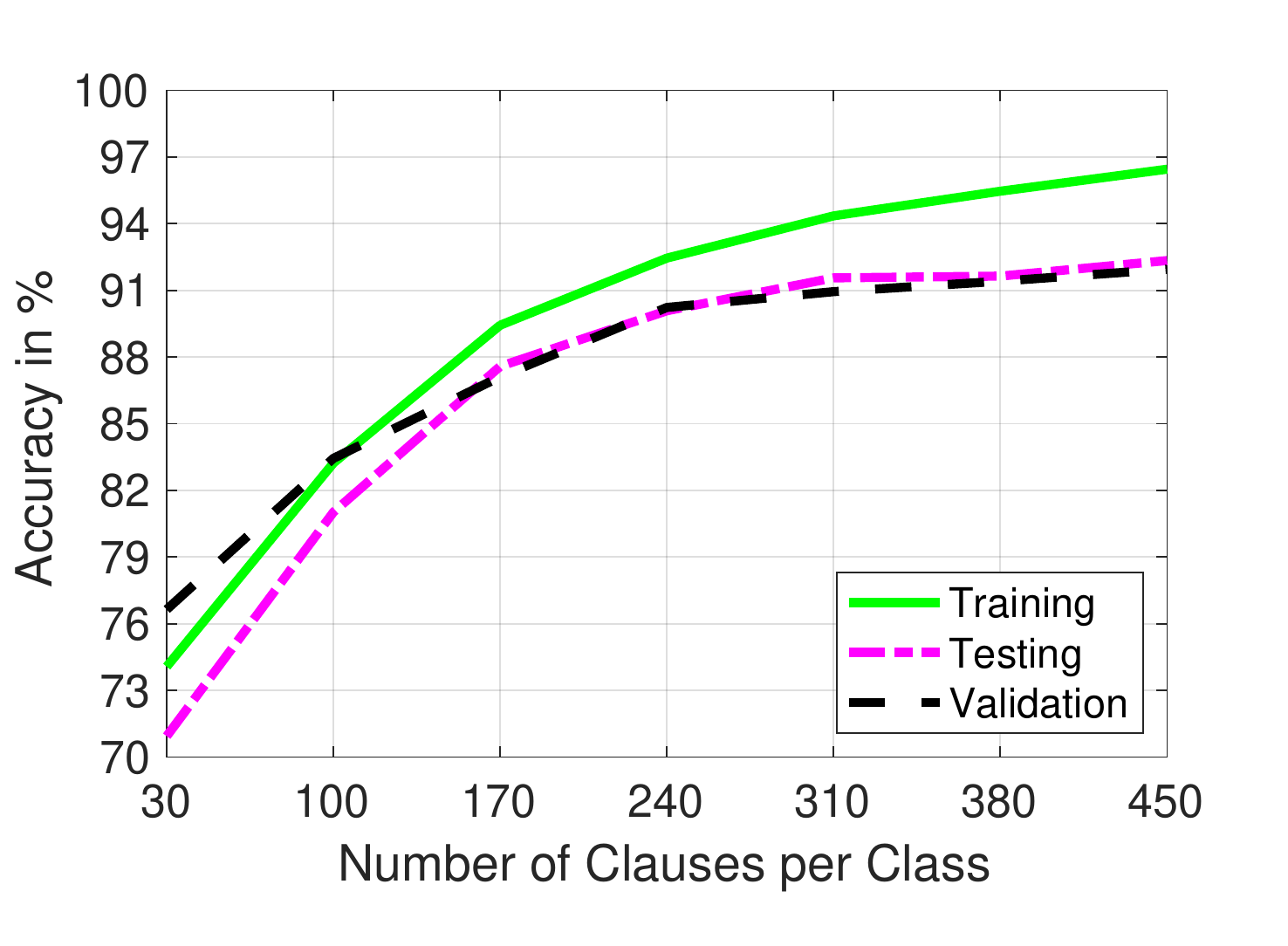} &     \includegraphics[width=0.45\columnwidth]{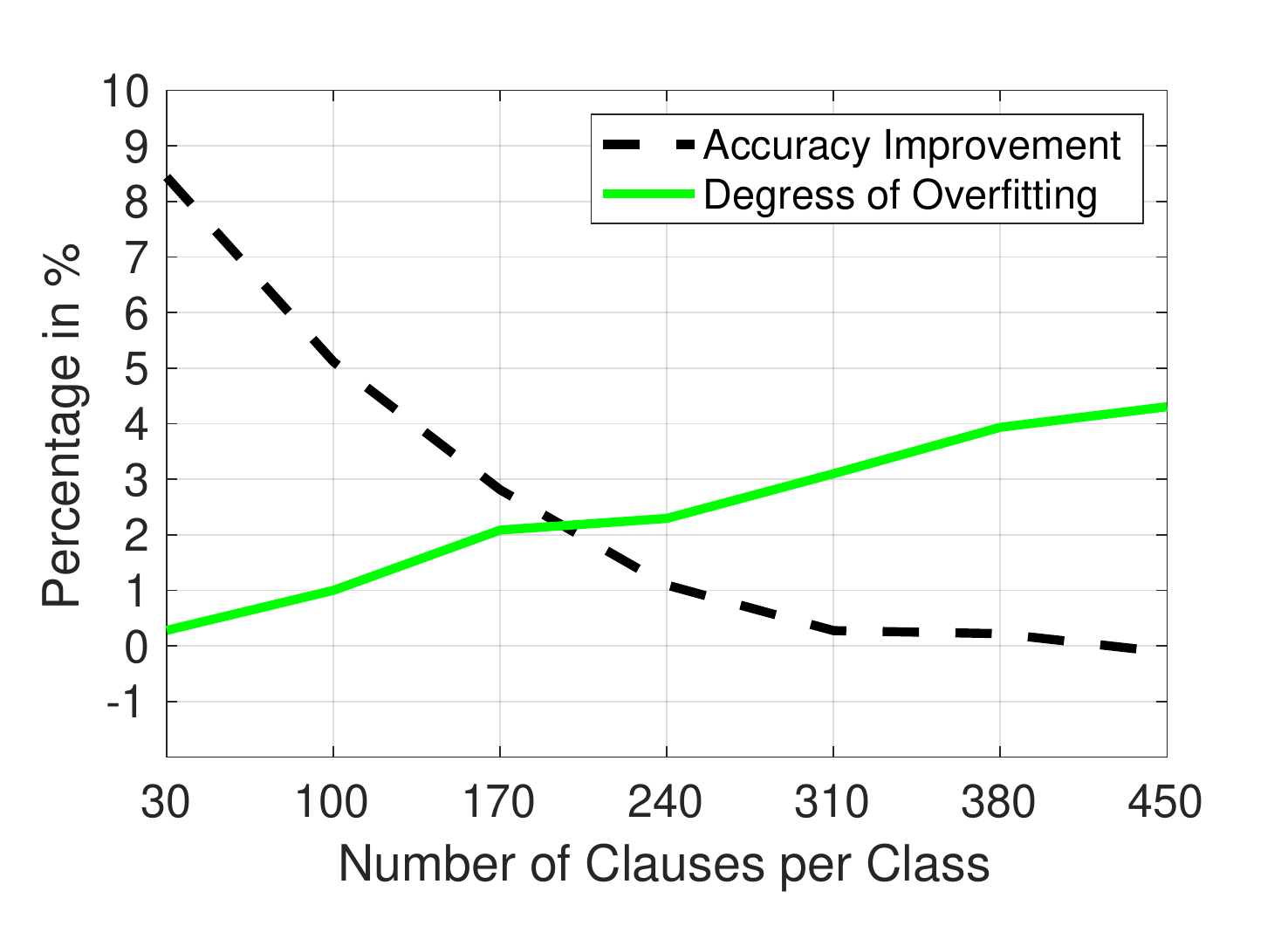}\\
         \small{(a) The effect on accuracy.} &
         \small{(b) The effect on overfitting.}
        \end{tabular}
        \caption{Effect of increasing the number of clauses on accuracy and overfitting.}
        \label{fig:wsp_sweepClause_all}
    \end{figure}

The influence of increasing the number of clauses was briefly discussed in Section \ref{sec:wsp_design_tm}, here we can see the experimental result in Figure \ref{fig:wsp_sweepClause_all}(a) showing the impact of increasing clauses with 4 classes. 
    
Increased number of clauses leads to better performance. However, upon closer examination we can also see the impact of overfitting at the clause level, i.e., increasing the number of clauses has resulted in a larger difference in the training accuracy with the testing and validation. The datapoints for the 4 classes were sufficient to create largely different sub-patterns for the TAs during training, but not complete enough to describe new data in the testing and validation. 
    
\begin{figure}[ht] 
    \begin{tabular}{ c c }
    \centering
         \includegraphics[width=0.45\columnwidth]{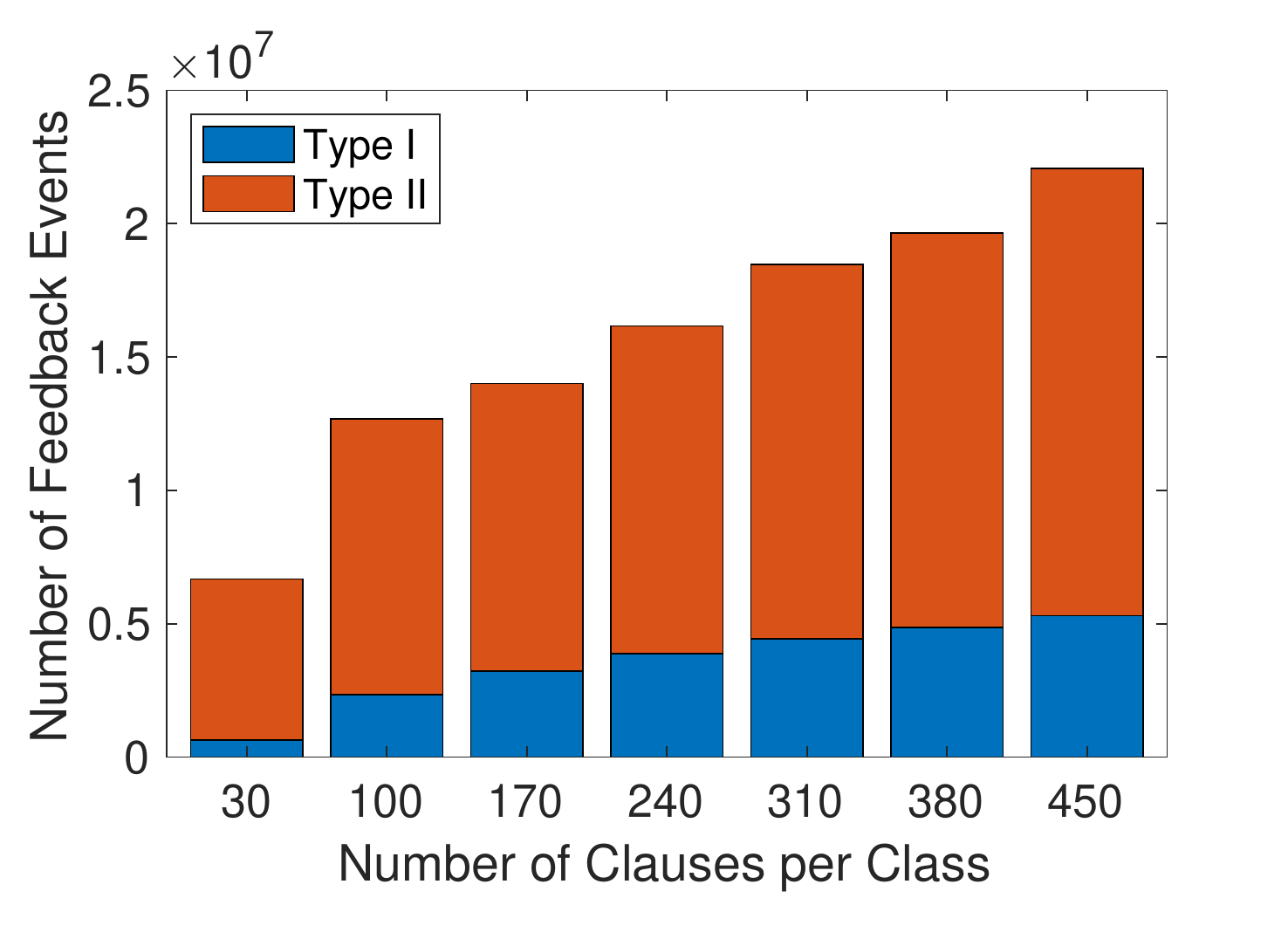} &     \includegraphics[width=0.45\columnwidth]{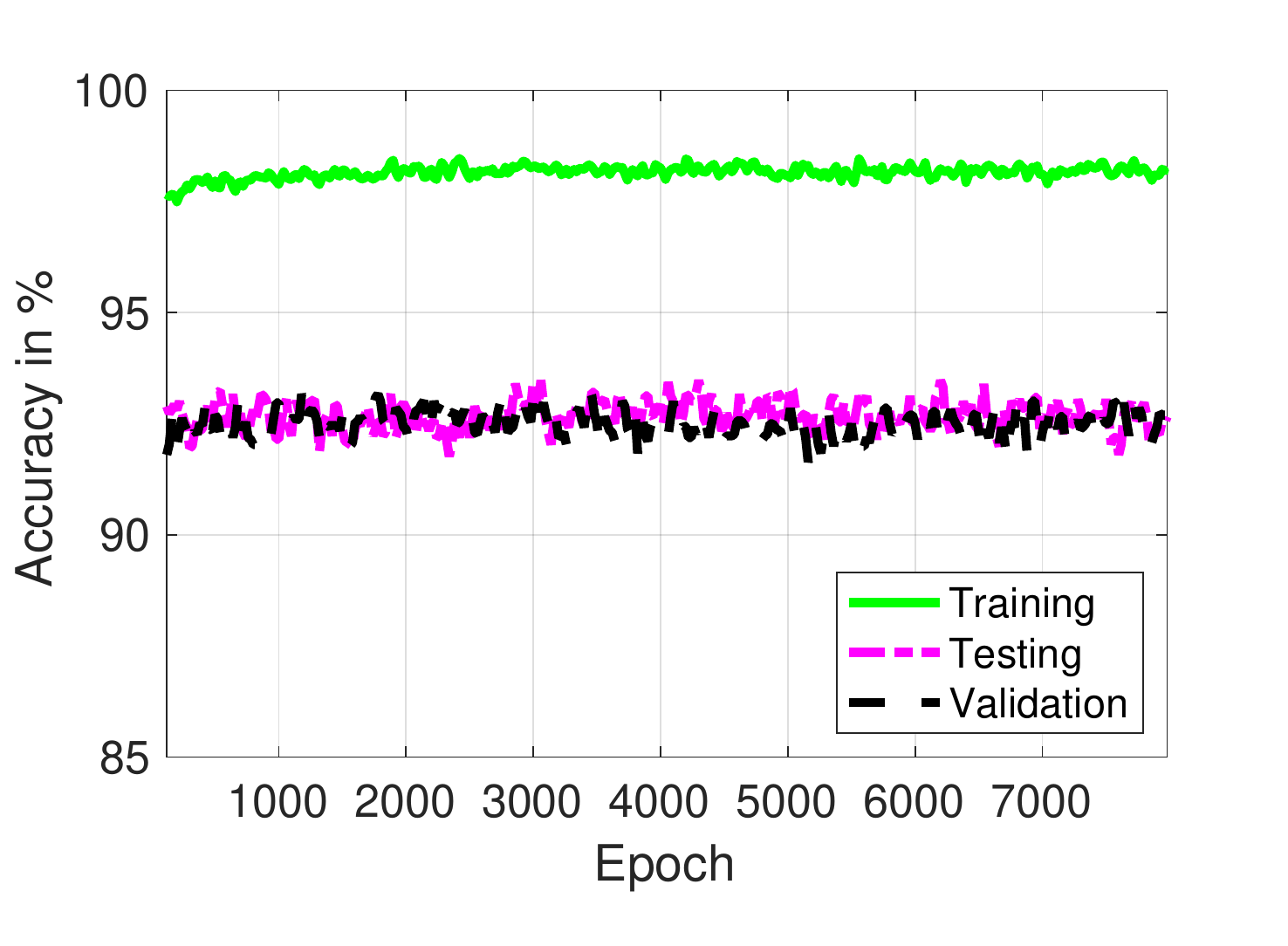}\\
         \small{(a) The effect of clauses on feedback.} &
         \small{(b) The effect of epoch on accuracy.}
        \end{tabular}
        \caption{Effect of increasing the number of clauses on \ac{TM} feedback (a), and the effect of increasing the number of epochs on accuracy (b).}
        \label{fig:wsp_manyClauseDisscu_all}
    \end{figure}

As a result, when clauses are increased, more clauses reach incorrect decisions and sway the voting in the summation and threshold module toward incorrect classification, which is seen through Figure \ref{fig:wsp_manyClauseDisscu_all}(a). The \ac{TM} has two types of feedback, Type I, which introduces stochasticity to the system and Type II, which bases state transitions on the results of corresponding clause value. Type II feedback is predominantly used to diminish the effect of false positives. We see that as the clause value increases the \ac{TM} uses more Type II feedback indicating increased false positive classifications. This result is for due to the incompleteness in the training data in describing all possible logic propositions for each class. We see this through \ref{fig:wsp_manyClauseDisscu_all}(b); despite increasing the number of epochs we do not experience a boost in testing and validation accuracy and through Figure \ref{fig:wsp_sweepClause_all}(b) we find the point where the overfitting outweighs the accuracy improvement at around 190-200 clauses.

From the perspective of energy efficiency, these results offer two possible implications for the \ac{KWS}-\ac{TM} pipeline, if a small degradation of performance in the \ac{KWS} application is acceptable, then operating at a lower clause range will be more beneficial for the \ac{TM}. The performance can then be boosted through hyperparameters available to adjust feedback fluidity. This approach will reduce energy expenditure through fewer clause computations and reduce the effects of overfitting when the training data lacks enough completeness. Alternatively, if performance is the main goal, then the design focus should be on injecting training data with more diverse datapoints to increase the descriptiveness of each class. In that case, increased clauses will provide more robust functionality.  
    
        \begin{table}[htbp]
        \caption{Impact of the number of clauses on energy\slash accuracy tradeoffs.}
        \centering
        \begin{tabular}{c||c|c|c|c|c}
                 & Clauses & Current & Time & Energy  & Accuracy\\
        \hline
        \hline
        Training & 100 & 0.50 A  & 68 s   &  426.40 J  &- \\
        Training & 240 & 0.53 A  & 96 s   &  636.97 J &-    \\
        \hline
        Inference   & 100  & 0.43 A  &  12 s  &   25.57 J  & 80 \% \\
        Inference   & 240  & 0.47 A  &  37 s  &  87.23 J & 90 \%\\
        \hline
        \end{tabular}
                \newline
        \label{tab:wsp_clauseEnergy}
    \end{table}
    
The impacts of being resource efficient and energy frugal are most prevalent when implementing \ac{KWS} applications into dedicated hardware and embedded systems. To explore this practically, the \ac{KWS}-\ac{TM} pipeline was implemented onto a Raspberry Pi. The same 4 keyword experiement was ran with 100 and 240 clauses. As expected, we see that increased clause computations lead to increased current, time and energy usage, but also delivers better performance. We can potentially boost the performance of the Tsetlin Machine at lower clauses through manipulating the hyperparameters as seen Table \ref{tab:wsp_clauseT}. 

\begin{table}[htbp]
        \caption{Impact of the \textit{T} values on accuracy }
        \centering
        \begin{tabular}{c||c|c|c|c|c}
         Clauses     & T & Training & Testing & Validation & Better Classification \\
        \hline
        \hline
        30 & 2 & 83.5 \%  & 80.5 \%   &  83.8 \%  & \checkmark\\
        30 & 23 & 74.9 \%  & 71.1 \%   &  76.1 \%  & \\
        \hline
        450 & 2 & 89.7 \%  & 86.1 \%   &  84.9 \%  & \\
        450 & 23 & 96.8 \%  & 92.5 \%   &  92.7 \%  & \checkmark\\
        \hline
        \end{tabular}
                \newline
        \label{tab:wsp_clauseT}
\end{table}
    
The major factor that has impacted the performance of the \ac{KWS} is the capacity of the \ac{TM} which is determined by the number of clauses per class. The higher the number clauses, the higher the overall classification accuracy \cite{Wheeldon2020a}. Yet, the resource usage will increase linearly along with the energy consumption and memory footprint. Through Table \ref{tab:wsp_clauseT} we see that at 30 clauses the accuracy can be boosted through reducing the Threshold hyperparameter. The table offers two design scenarios; firstly, very high accuracy is achievable through a large number of clauses (450 in this case) and a large Threshold value. With a large number of clauses an increased number of events must be triggered in terms of state transitions (see Figure \ref{fig:tm_st_stoca}) to encourage more feedback to clauses and increases the \ac{TM}s decisiveness. While this offers a very good return on performance, the amount of computations are increased with more clauses and more events triggered and this leads to increased energy expenditure as seen through Table \ref{tab:wsp_clauseEnergy}. 
    
In contrast, using 960 clauses and a lower Threshold still yields good accuracy but at a much lower energy expenditure through fewer clause computations and feedback events. A smaller number of clauses mean that the vote of each clause has more impact, even at a smaller Threshold the inbuilt stochasticity of the \ac{TM}'s feedback module allows the TAs to reach the correct propositional logic. Through these attributes it is possible to create more energy frugal \ac{TM}s requiring fewer computations and operating at a much lower latency.

\subsection{Comparative Learning Convergence and Complexity Analysis of KWS-TM} \label{sec:comparative}

Both \acp{TM} and \acp{NN} have modular design components in their architecture; For the \ac{TM}, this is in the form of clauses and for the \ac{NN} it is the number of neurons. \acp{NN} require input weights for the learning mechanism which define the neurons' output patterns. The number of weights and the number of neurons are variable, however more neurons will lead to better overall \ac{NN} connectivity due to more refined arithmetic pathways to define a learning problem. 

For the \ac{TM} the clauses are composed of TAs. The number of TAs are defined by the number of Boolean features which remains static throughout the course of the \acp{TM} learning. It is the number of clauses that is variable, increasing the clauses typically offers more propositional diversity to define a learning problem.

\begin{figure}[ht] 
    \begin{tabular}{ c c }
    \centering
         \includegraphics[width=0.45\columnwidth]{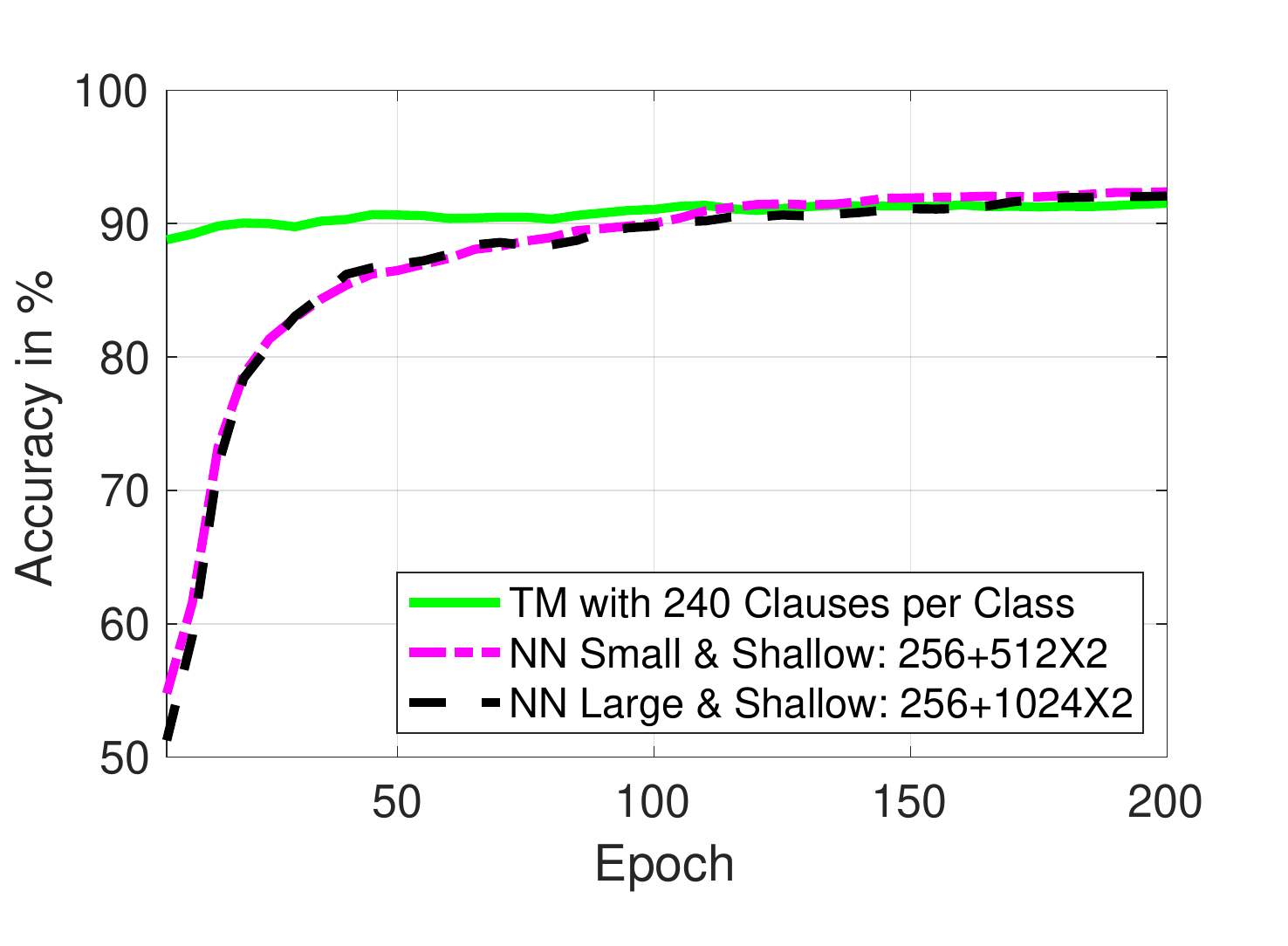} &     \includegraphics[width=0.45\columnwidth]{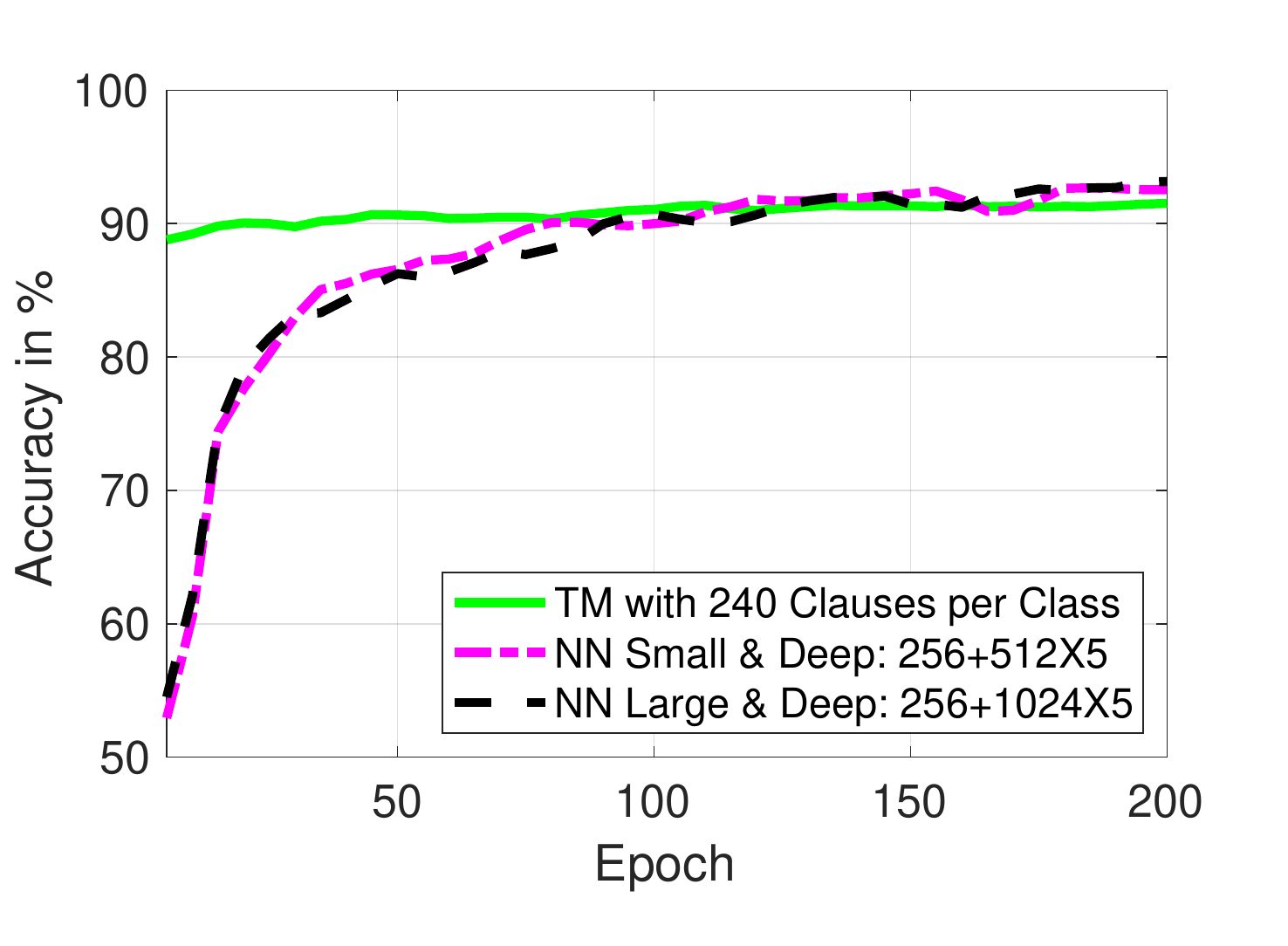}\\
         \small{(a) Convergence of the TM against shallow NNs.} &
         \small{(b) Convergence of the TM against deep NNs.}
        \end{tabular}
        \caption{Training convergence of TM and NN implementations.}
        \label{fig:wsp_NNcompTM_kws_all}
    \end{figure}

Through Figure \ref{fig:wsp_NNcompTM_kws_all} and Table \ref{tab:wsp_nn_tm_parameters_9keykws} we investigate the learning convergence rates of the \ac{TM} against 4 'vanilla' \ac{NN} implementations. The \ac{TM} is able to converge to 90.5$\%$ after less than 10 epochs highlighting its quick learning rate compared to \ac{NN}s which require around 100 epochs to converge to the isoaccuracy target ($\approx$90\%). After further 100 epochs the \ac{NN} implementations reach only marginally better accuracy than \ac{TM}. The indirect implication of faster convergence is improved energy efficiency as fewer training epochs will result in fewer computations required for the TA states to settle. 

Table \ref{tab:wsp_nn_tm_parameters_9keykws} shows one of the key advantages of the \ac{TM} over all types of \ac{NN}s, the significantly fewer parameters required, i.e. low-complexity. Large number of parameters needed for \ac{NN}s are known to limit their practicality for on-chip \ac{KWS} solutions \cite{Zhang2018, CNN, 8936893, 8502309}, where as the \ac{TM} offers a more resource-frugal alternative. With only 960 clauses, which require only logic based processing, the \ac{TM} outperforms even the most capable large and deep \ac{NN}s. In our future work, we aim to exploit this to enable on-chip learning based KWS solutions.

    \begin{table}[htbp]
        \caption{The required parameters for different \ac{NN}s and the \ac{TM} for a 4 keyword problem.}
        \centering
        \begin{tabular}{p{2.3in}||p{1in}|p{1.45in}}
                KWS-ML Configuration & Num. neurons & Num. hyperparameters  \\
        \hline
        \hline
        NN Small \& Shallow: 256+512X2 & 1,280  & 983,552    \\
        NN Small \& Deep: 256+512X5   & 2,816  & 2,029,064    \\
        NN Large \& Shallow: 256+1024X2   & 2,304  &  2,822,656  \\
        NN Large \& Deep: 256+1024X5   & 5,376  & 7,010,824  \\
        \hline
        TM with 240 Clauses per Class   & 960 (clauses) & 2 hyperparameters with 725760 TAs \\
        \hline
        \end{tabular}
                \newline
        \label{tab:wsp_nn_tm_parameters_9keykws}
    \end{table}
	\section{Related Work}
This section will provide a brief examination into current \ac{KWS} research, industrial challenges with \ac{KWS}, deeper look in the component blocks of the \ac{TM} and provide insight into the current developments and the future research directions. 

\subsection{Current \ac{KWS} developments} \label{sec:RelatedKWS} 

The first \ac{KWS} classification methods proposed in the late 1970s used \ac{MFCC}s for their feature extraction ability and because the coefficients produced offered a very small dimensionality compared to the raw input data that was being considered then \cite{Wilpon1990}. It was later shown that compared to other audio extraction methods such as near prediction coding coefficients (LPCC)s and perceptual linear production (PLP), \ac{MFCC}s perform much better with increased background noise and low SNR \cite{8936893}. 

For the classifier, Hidden Markov Models (HMMs) were favored after the \ac{MFCC} stage due to their effectiveness in modelling sequences \cite{Wilpon1990}. However they rely on many summation and Bayesian probability based arithmetic operations as well as the computationally intensive \emph{Viterbi} decoding to identify the final keyword \cite{Zhang2018, RNNs, Chen2014}.

Later it was shown that Recurrent Neural Networks (RNN)s outperform HMMs but suffer from operational latency as the problem scales, albeit RNNs still have faster run-times than HMM pipelines given they do not require a decoder algorithm \cite{RNNs}. To solve the latency issue, the \ac{DNN} was used, it has smaller memory footprint and reduced run-times compared to HMMs \cite{8936893,Chen2014}. However, \ac{DNN}s are unable to efficiently model the temporal correlations of the \ac{MFCC}s and their transitional variance \cite{CNN} \cite{Zhang2018}. In addition to this, commonly used optimization techniques used for \ac{DNN}s such as pruning, encoding and quantization lead to great accuracy losses with \ac{KWS} applications \cite{8936893}.   

The \ac{MFCC} features exist as a 2D array as seen in Figure \ref{fig:wsp_mfcc_tm}, to preserve the temporal correlations and transitional variance, this array can be treated as an image and a \ac{CNN} can be used for classification \cite{8502309, CNN}. With the use of convolution comes the preservation of the spatial and temporal dependencies of the 2D data as well as the reduction of features and computations from the convolution and pooling stages \cite{8502309}. However, once again both the \ac{CNN} and \ac{DNN} suffer from the large number of parameters (250K for the dataset used in \cite{CNN} and 9 million Multiplies required for the \ac{CNN}). Despite the gains in performance and reductions in latency, the computational complexity and large memory requirements from parameter storage are ever present with all \ac{NN} based \ac{KWS} solutions.

The storage and memory requirements played a major part in transitioning to a micro-controller system for inference where memory is limited through the size of the SRAM \cite{Zhang2018}. In order to accommodate for the large throughput of running \ac{NN} workloads, micro-controllers with integrated DSP instructions or integrated SIMD and MAC instructions can accelerate low-precision computations \cite{Zhang2018}. When testing for 10 keywords, it was shown experimentally in \cite{Zhang2018}, that for systems with limited memory and compute abilities \ac{DNN}s are favorable given they use the fewer operations despite having a lower accuracy (around 6$\%$ less) compared to \ac{CNN}s.

It is when transitioning to hardware that the limitations of memory and compute resources become more apparent. In these cases it is better to settle for energy efficiency through classifiers with lower memory requirements and operations per second even if there is a slight drop in performance.

A 22nm CMOS based \ac{QCNN} Always-ON \ac{KWS} accelerator is implemented in \cite{8936893}, they explore the practicalities of \ac{CNN} in hardware through quantized weights, activation values and approximate compute units. Their findings illustrate the effectiveness of hardware design techniques; the use of approximate compute units led to a significant decrease in energy expenditure, the hardware unit is able to classify 10 real-time keywords under different SNRs with a power consumption of 52$\mu$W. This impact of approximate computing is also argued in \cite{8502309} with design focus given to adder design, they propose an adder with a critical path that is 49.28$\%$ shorter than standard 16-bit Ripple Carry Adders.

Through their research work with earables Nokia Bell Labs Cambridge have brought an industrial perspective to the idea of functionality while maintaining energy frugality into design focus for \ac{AI} powered \ac{KWS} \cite{10.1145/3211960.3211970,8490189}, with particular emphasis on user oriented ergonomics and commercial form factor. They discovered that earable devices are not as influenced by background noise compared to smartphones and smartwatches and offer better signal-to-noise ratio for moving artefacts due to their largely fixed wearing position in daily activities (e.g. walking or descending stairs) \cite{8490189}. This was confirmed when testing using Random Forest classifiers.

\subsection{The Tsetlin Machine}\label{sec:RelatedTM}

\begin{figure}[htbp]
    \centering
    \includegraphics[width=0.7\columnwidth]{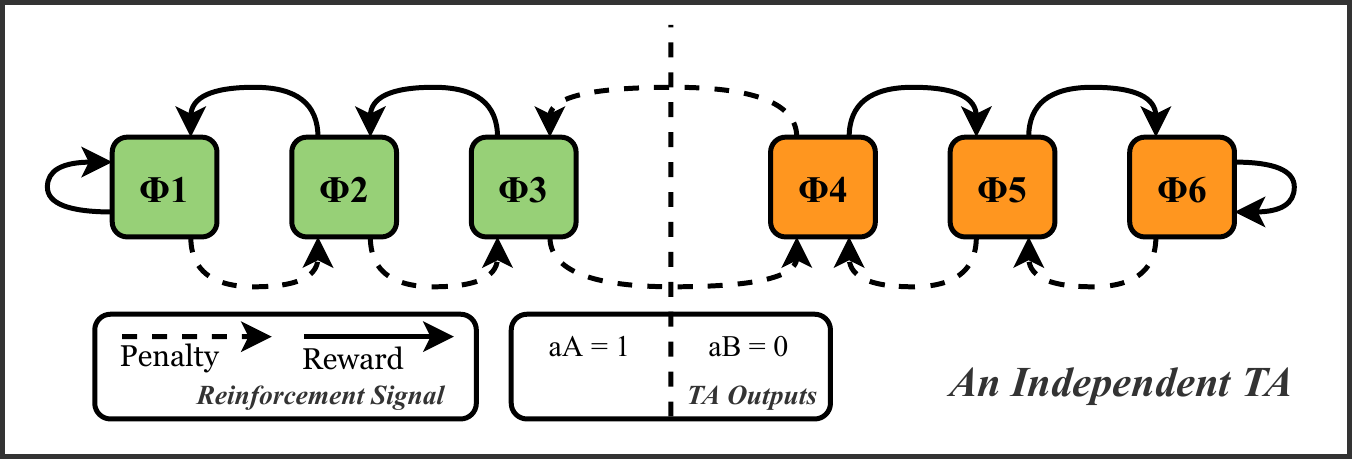}
    \caption{Mechanism of a TA.}
    \label{fig:wsp_ta}
\end{figure}

\begin{figure}[htbp]
    \centering
    \includegraphics[width=0.8\columnwidth]{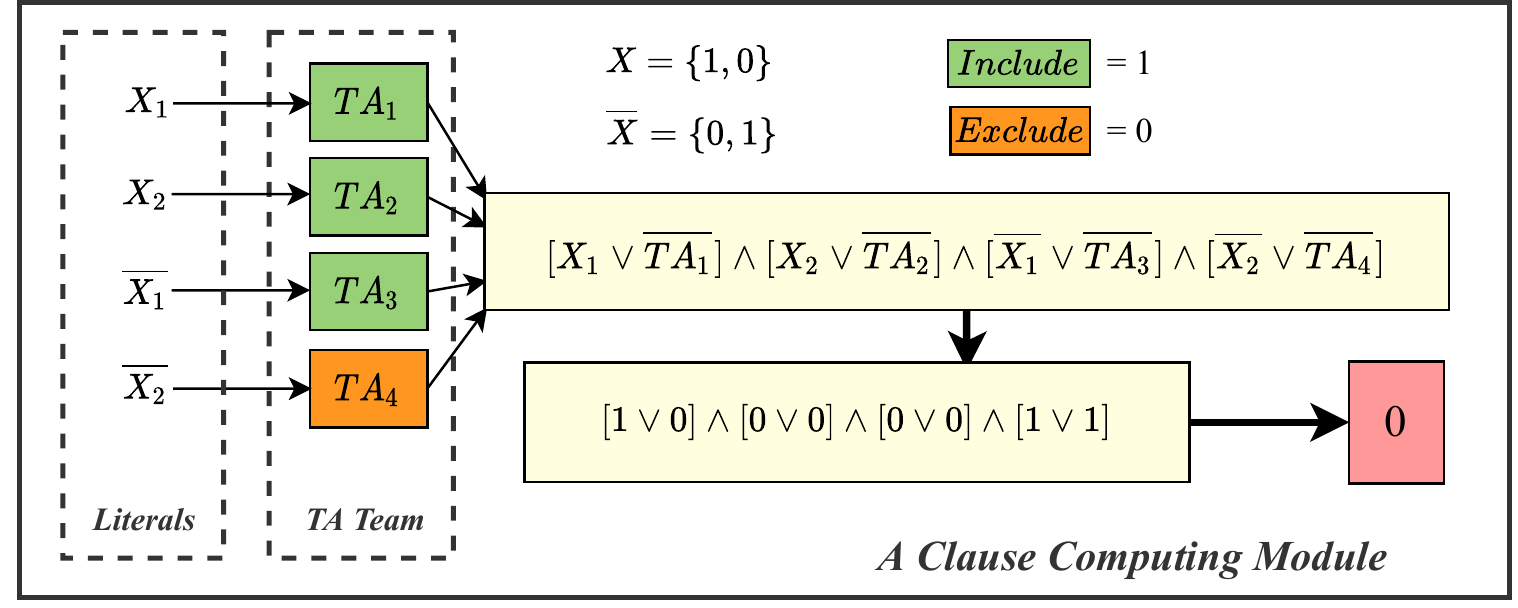}
    \caption{Mechanism of a Clause computing module (assuming TA\textsubscript{1}= 1 means \emph{Include} and TA\textsubscript{1}= 0 means \emph{Exclude}).}
    \label{fig:wsp_clause}
\end{figure}

\begin{figure}[htbp]
    \centering
    \includegraphics[width=0.5\columnwidth]{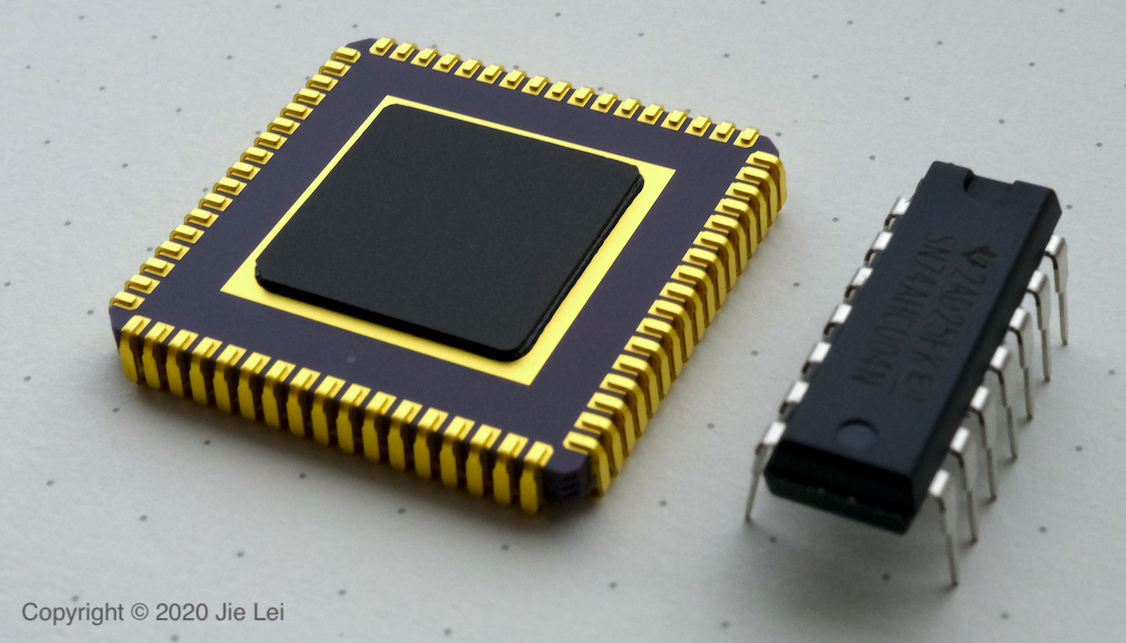}
    \caption{\href{www.mignon.ai}{The Mignon AI}: ASIC microchip acclerating \ac{TM} (left).}
    \label{fig:asic}
\end{figure}

We briefly discussed the overall mechanism of the \ac{TM} and the main building blocks in the \cref{sec:wsp_design_tm}. In this section, we will have a closer look to the fundamental learning element of the \ac{TM}, namely the Tsetlin Automaton, as described in \cref{fig:wsp_ta}. We will also present a more detailed look at the clause computing module as seen in \cref{fig:wsp_clause}, and we will discuss the first \ac{ASIC} implementation of the \ac{TM}, the Mignon \footnote{ Mignon AI: http://mignon.ai/}, as seen in  \cref{fig:asic}.

The \ac{TA} is the most fundamental part of the \ac{TM} forming the core learning element that drives classification (\cref{fig:wsp_ta}). Developed by Mikhail Tsetlin in the 1950s, the \ac{TA} is an \ac{FSM} where the current state will transition towards or away from the  middle state upon receiving \textit{Reward} or \textit{Penalty} reinforcements during the \ac{TM}s training stage. The current state of the \ac{TA} will decide the output of the automaton which will be either an  \textit{Include} (aA) or \textit{Exclude} (aB). 

 \cref{fig:wsp_clause} shows how the clause module create logic propositions that describe the literals based on the TA decisions through logic \emph{OR} operations between the negated TA decision and the literal. The TA decision is used to bit mask the literal and through this we can determine which literals are to be excluded. The proposition is then logic \emph{AND}ed and this forms the raw vote for this clause. Clauses can be of positive and negative polarity, as such, a sign will be added to the clause output before it partakes in the class voting. It is important to note the reliance purely on logic operations making the \ac{TM} well suited to hardware implementations. Clauses are largely independent of each other, only coalescing for voting giving the \ac{TM} good scalability potential.
 
The feedback to the \ac{TM} can be thought of on three levels, at the \ac{TM} level, at the clause level and at the TA level. At the \ac{TM} level, the type of feedback to issue is decided based on the target class and whether the \ac{TM} is in learning or inference mode. For inference no feedback is given, we simply take the clause computes for each class and pass to the summation and threshold module to generate the predicted class. However, in training mode there is a choice of Type I feedback to combat false negatives or Type II feedback to combat false positives. This feedback choice is further considered at the clause level. 

At the clause level there are three main factors that will determine feedback type to the TAs, the feedback type decision from the TM level, the current clause value, and whether the magnitude of clause vote is above the magnitude of the Threshold. 

At the TA level, the feedback type from the clause level will be used in conjunction with the current TA state and the s parameter to determine whether there is inaction, penalty or reward given to the TA states. 

The simplicity of the \ac{TM} shows its potential to be a promising \ac{NN} alternative. Lei \textit{ et al } \cite{jieleitmnn} comparatively analyzed the architecture, memory footprint and convergence of these two algorithms for different datasets.
This research shows the fewer number of hyperparameter of the \ac{TM} will reduce the complexity of the design. The convergence of the \ac{TM} is higher than the \ac{NN} in all experiments conducted. 

The most unique architectural advances of the \ac{TM} is the propositional logic based learning mechanism which will be beneficial in achieving energy frugal hardware \ac{AI}. Wheeldon \textit{et al.} \cite{Wheeldon2020a} presented the first \ac{ASIC} implementation of the \ac{TM} for Iris flower classifications (see \cref{fig:asic}). 

This 65-nm technology based design is a breakthrough in achieving an energy efficiency of up to 63 Tera Operations per Joule (Tops\slash Joule) while maintaining high convergence rate and performance. The early results from this microchip has been extensively compared with \ac{BCNN} and neuromorphic designs in \cite{Wheeldon2020a}. 

In addition, Wheeldon \textit{et al.} \cite{Wheeldon2020a} also proposed a system-wide design space exploration pipeline in deploying \ac{TM} into \ac{ASIC} design. They introduced a detailed methodology from 1) dataset encoding building on the work seen in \cite{9308291} to 2) software based design exploration and 3) an FPGA based hyperparameter search to 4) final \ac{ASIC} synthesis. A follow-up work of this~\cite{wheeldon2020low} also implemented a self-timed and event-driven hardware TM. This implementation showed power and timing elasticity properties suitable for low-end AI implementations at-the-microedge.

Other works include mathematical lemma based analysis of clause convergence using the XOR dataset \cite{jiao2021convergence}, natural language (text) processing \cite{bhattarai2020}, disease control \cite{9308291}, methods of automating the \textit{s} parameter \cite{gorji2019tsetlin} as well as exploration of regression and convolutional \ac{TM}s \cite{Abeyrathna, ConvTM}. 

 The \ac{TM} has so far, been implemented with many different programming languages such as, C, C++, C$\#$, Python and Node.js, to name a few. It has also been optimized for \ac{HPC} through \ac{CUDA} for accelerating \ac{GPU} based solutions and currently through OpenCL for heterogeneous embedded systems~\cite{abeyrathna2020massively}. 
 
 Exploiting the natural logic underpinning there are currently ongoing efforts in establishing explainability evaluation and analysis of TMs\cite{shafikexplainability}. Deterministic implementation of clause selection in TM, reported by~\cite{abeyrathna2020novel}, is a promising direction to this end.
 
 Besides published works, there are numerous talks, tutorials and multimedia resources currently available online to mobilize the hardware\slash software community around this emerging \ac{AI} algorithm. Below are some key sources:

Videos: \textit{\url{https://tinyurl.com/TMVIDEOSCAIR}}.

Publications: \textit{\url{https://tinyurl.com/TMPAPERCAIR}} \& \textit{\url{www.async.org.uk}}.

Software implementations: \textit{\url{https://tinyurl.com/TMSWCAIR}}

Hardware implementations, Mignon AI: \textit{\url{http://www.mignon.ai/}}.

A short video demonstrating \ac{KWS} using \ac{TM} can be found here:\\ \indent \textit{\url{https://tinyurl.com/KWSTMDEMO}}.

	\section{Summary and Conclusions} \label{sec:conclusions}

The paper presented the first ever \ac{TM} based \ac{KWS} application. Through experimenting with the hyperparameters of the proposed \ac{KWS}-\ac{TM} pipeline we established relationships between the different component blocks that can be exploited to bring about increased energy efficiency while maintaining high learning efficacy.

From current research work we have already determined the best methods to optimize for the \ac{TM} is through finding the right balance between reduction of the number of features, number of clauses and number of events triggered through the feedback hyper-parameters against the resulting performance from these changes. These insights were carried into our pipeline design exploration experiments.  

Firstly, we fine tuned the window function in the generation of \ac{MFCC}s, we saw that increasing the window steps lead to much fewer \ac{MFCC}s and if the window length is sufficient enough to reduce edge tapering then the performance degradation is minimal. Through quantile binning to manipulate the discretization of the Boolean \ac{MFCC}s, it was seen that this did not yield change in performance. The \ac{MFCC} features of interest have very large variances in each feature column and as such less precision can be afforded to them, even as low as one Boolean per feature. This was extremely useful in reducing the resulting \ac{TM} size. 

Through manipulating the number of clause units to the \ac{TM} on a Raspberry Pi, we confirmed the energy and latency savings possible by running the pipeline at a lower clause number and using the Threshold hyper-parameter the classification of the accuracy can also be boosted. Through these design considerations we are able to increase the energy frugality of the whole system and transition toward low-power hardware accelerators of the pipeline to tackle real time applications.  

The \ac{KWS}-\ac{TM} pipeline was then compared against some different \ac{NN} implementations, we demonstrated the much faster convergence to the same accuracy during training. Through these comparisons we also highlighted the far fewer parameters required for the \ac{TM} as well as a fewer number of clauses compared to neurons. The faster convergence, fewer parameters and logic over arithmetic processing makes the \ac{KWS}-\ac{TM} pipeline more energy efficient and enables future work into hardware accelerators to enable better performance and low power on-chip \ac{KWS}.

\textbf{Acknowledgement}: The authors gratefully acknowledge the funding from EPSRC IAA project ``Whisperable'' and EPSRC grant STRATA (EP/N023641/1). The research also received help from the computational powerhouse at CAIR\footnote{https://cair.uia.no/house-of-cair/}.
    \section{Future Work}\label{sec:future}
Through testing the \ac{KWS}-\ac{TM} pipeline against the Tensorflow Speech data set we did not account for background noise effects. In-field \ac{IoT} applications must be robust enough to minimize the effects of additional noise, therefore, future work in this direction should examine the effects of the pipeline with changing signal-to-noise ratios. The pipeline will also be deployed to a micro-controller in order to benefit from the effects of energy frugality by operating at a lower power level.

\bibliographystyle{unsrt}  

\bibliography{tsetlin,audioTM}





\end{document}